\documentclass[useAMS,usegraphicx,usenatbib]{mn2e}
\usepackage{amsbsy,amsmath,amssymb,amstext,longtable,times,xspace}
\usepackage{natbib}

\title[Line-depth ratios for $\nu$ Octantis]{Line-depth-ratio temperatures for 
the close binary \mbox{$\nu$ Octantis}: new evidence supporting the  
conjectured retrograde planet}

\author[D.~J.~Ramm]{D.~J.~Ramm\thanks{E-mail: djr1817@gmail.com} \\
Department of Physics and Astronomy, University of Canterbury, 
Christchurch 8020, New Zealand \\
}

\newcommand*{\nc}{\newcommand*}

\nc{\esm}{\ensuremath}
\nc{\xs}{\xspace}
\nc{\oo}{et~al.\xs}

\nc{\dr}{Ramm \oo\xs}
\nc{\drb}{Ramm \oo (2009)\xs}
\nc{\dry}{(Ramm \oo 2009)\xs}
\nc{\nuo}{\esm{\nu~{\rm Oct}}\xs}
\nc{\se}{\section}
\nc{\su}{\subsection}
\nc{\suu}{\subsubsection}
\nc{\noi}{\\ \noindent}
\nc{\sci}{{\,\scriptsize I}\normalsize~}
\nc{\scii}{{\,\scriptsize II}\normalsize~}

\nc{\sss}{\rm \scriptscriptstyle}
\nc{\lab}{\label}

\nc{\ml}{\multicolumn}
\nc{\cpp}{\citep}
\nc{\cyp}{\citet}

\nc{\wide}{\hspace{1em}}
\nc{\es}{\hspace{1em}}
\nc{\dgap}{\hspace{2em}}
\nc{\tg}{\hspace{3em}}
\nc{\ngap}{\hspace{-2em}}

\nc{\mc}{\mathcal}
\nc{\mjup}{\esm{\mc{M}_{\rm Jup}}\xs}
\nc{\ppert}{\esm{P_{\sss pert}}\xs}
\nc{\rad}{\esm{\mc R}\xs}
\nc{\mass}{\esm{\mc M}\xs}
\nc{\radp}{\esm{\mc{R}_1}\xs}
\nc{\rads}{\esm{\mc{R}_2}\xs}
\nc{\teff}{\esm{T_{\rm eff}}\xs}
\nc{\msun}{\esm{~\mc{M}_{\odot}}\xs}
\nc{\lsun}{\esm{~\mc{L}_{\odot}}\xs}
\nc{\rsun}{\esm{~\mc{R}_{\odot}}\xs}
\nc{\mbol}{\esm{M_{\rm bol}}\xs}
\nc{\mbv}{\esm{M_{\rm V}}\xs}
\nc{\ang}{\esm{\rm \AA}\xs}
\nc{\sqrchi}{\esm{(\redchi)^{1/2}}\xs}
\nc{\redchi}{\esm{\chi ^2_{\sss \nu}}\xs}
\nc{\ld}{\esm{\lambda}\xs}
\nc{\abin}{\esm{a_{_{\rm bin}}}\xs}
\nc{\apl}{\esm{a_{_{\rm pl}}}\xs}
\nc{\ppl}{\esm{P_{_{\rm pl}}}\xs}
\nc{\pp}{^{\prime\prime}}

\nc{\f}{\frac}
\nc{\kone}{\esm{\rm 1k\times1k\xs}}
\nc{\kfour}{\esm{\rm 4k\times4k\xs}}
\nc{\sigc}{\esm{S_{\rm c}}\xs}
\nc{\sigp}{\esm{S_{\rm p}}\xs}

\nc{\mdiff}{\esm{\Delta M_{\rm\sss (V-ZAMS)}}\xs}
\nc{\tbv}{\esm{T_{(B-V)_{_0}}\xs}}
\nc{\vep}{\esm{\varepsilon}\xs}
\nc{\ldr}{\esm{_{_{\rm LDR}}}\xs}
\nc{\mldr}{\esm{_{_{\rm MLDR}}}\xs}
\nc{\prot}{\esm{P_{_{\rm rot}}}\xs}
\nc{\delt}{\esm{\delta T}\xs}
\nc{\Delt}{\esm{\Delta T}\xs}
\nc{\pc}{\esm{p_{_{\rm c}}}\xs}
\nc{\kpul}{\esm{K_{_{\rm ps}}}\xs}
\nc{\krv}{\esm{K_{_{\rm RV}}}\xs}

\nc{\om}{\esm{\omega}\xs}
\nc{\dg}{\esm{\degr}\xs}

\nc{\kms}{\esm{~\rm km\,s^{-1}}\xs}
\nc{\ms}{\esm{~\rm m\,s^{-1}}\xs}
\nc{\skms}{\esm{\,\rm km\,s^{\sss -1}}\xs}
\nc{\sms}{\esm{\,\rm m\,s^{\sss -1}}\xs}
\nc{\hc}{HERCULES\xs}

\nc{\hip}{{\it Hipparcos}\xs}

\nc{\ie}{i.e.~}
\nc{\eg}{e.g.~}
\nc{\fn}{\footnote}
\nc{\bfs}{\bfseries}

\nc{\bc}{\begin{center}}
\nc{\ec}{\end{center}}
\nc{\bte}{\begin{table}}
\nc{\ete}{\end{table}}
\nc{\btr}{\begin{tabular}}
\nc{\etr}{\end{tabular}}

\nc{\ds}{\displaystyle}
\nc{\beq}{\begin{equation}}
\nc{\eeq}{\end{equation}}
\nc{\bse}{\begin{subequations}}
\nc{\ese}{\end{subequations}}
\nc{\bea}{\begin{eqnarray}}
\nc{\eea}{\end{eqnarray}}

\nc{\bfi}{\begin{figure}}
\nc{\efi}{\end{figure}}
\nc{\ca}{\caption}
\nc{\eql}{Eq.~\eqref}
\nc{\fl}{Fig.~\ref}
\nc{\tl}{Table~\ref}
\nc{\fnl}{Footnote~\ref}
\nc{\scl}{\S~\ref}

\nc{\si}{\esm{\sigma}\xs}
\nc{\er}{\esm{\pm}\xs}

\begin{document}
\lab{firstpage}

\maketitle

\begin{abstract}
We explore the possibly that either starspots or pulsations are the cause of a 
periodic radial-velocity signal ($P\sim400$~days) from the K-giant binary 
$\nu$~Octantis ($P\sim1050$~days, $e\sim0.25$), alternatively conjectured to 
have a retrograde planet. Our study is based on temperatures derived from 22 
line-depth ratios (LDRs) for \nuo and twenty calibration stars. Empirical 
evidence and stability modelling provide unexpected support for the planet 
since other standard explanations (starspots, pulsations and 
additional stellar masses) each have credibility problems. However, the 
proposed system presents formidable challenges to planet-formation and 
stability theories: it has by far the smallest stellar separation of any 
claimed planet-harbouring binary ($\abin\sim2.6$~AU) and an equally 
unbelievable separation ratio ($\apl/\abin\sim0.5$), hence the necessity that 
the circumstellar orbit be retrograde.

The LDR analysis of 215 \nuo spectra acquired between 2001--2007, from which 
the RV perturbation was first revealed, have no significant periodicity at any 
frequency. The LDRs recover the original 21 stellar temperatures with an 
average accuracy of $45\pm25$~K. The 215 \nuo temperatures have a standard 
deviation of only 4.2~K. Assuming the host primary is not pulsating, the 
temperatures converted to magnitude differences strikingly mimic 
the very stable photometric \hip observations 15~years previously, implying 
the long-term stability of the star and demonstrating a novel use of LDRs as a 
photometric gauge. Our results provide substantial new evidence that 
conventional starspots and pulsations are unlikely causes of the RV 
perturbation. The controversial system deserves continued attention, including 
with higher resolving-power spectra for bisector and LDR analyses.
\end{abstract}


\se{Introduction}
\lab{intro}
The single-lined spectroscopic binary (SB1) $\nu$~Octantis (HD\,205478, 
HIP\,107089, HD\,8254; $P\sim1050$~days), having a slightly evolved early 
K-type primary, has been conjectured to have a so-far unique retrograde 
circumstellar planetary orbit associated with 
it based on radial velocity observations over several years (Ramm 2004; 
Ramm \oo 2009). Ramm \oo more or less discounted all other standard 
causes for the $\sim400$~day periodic signal including rotation modulation 
of surface features, pulsations, and a prograde orbit (which stability models 
indicate is rapidly unstable). Significant surface dynamics were not supported 
by their bisector 
analysis and \hip observations (ESA 1997) had already found \nuo to be 
particularly photometrically stable 
(see \tl{stellar}). \nuo has been found consistently to be inactive and 
without significant variability at all other studied spectral regions, for 
instance Ca\scii (Warner 1969), radio (Slee \oo 1989; 
Beasley, Stewart \& Carter 1992) and X-ray \cpp{hunsch}. Instrument- and 
data-reduction-related causes for the RV behaviour were discounted by 
Ramm \oo as their paper included a similar SB1, $\beta$~Reticuli (K2~III; 
Gray \oo 2006) frequently observed on the same nights whose RVs had no such 
anomalous behaviour. The proposal by Morais \& Correia 
(2012) that \nuo is actually a hierarchical 
triple system has merit given how frequently such systems are anticipated and 
observed (Tokovinen \oo 2006; Tokovinen, Hartung \& Hayward 2010). However, 
this scenario is challenged by 
the lack of observational support. In particular, their model predicts an 
apsidal precession rate of $\rm -0.86\dg/yr$ for the 
primary star's orbit. Yet, the orbital solution for the historical RVs (which 
date back to 1904--1924), re-derived in Ramm (2004) and provided in \drb, 
suggests there is no such 
change, since these RVs yield $\omega_{_1}=82\pm14\dg$ and the RVs from 
2001--2006 yield $\omega_{_1}=75.05\pm0.08\dg$ \dry. The approximate 90-year 
time interval between the two datasets corresponds to about 80\dg of predicted 
precession which is not at all apparent.

\bte
\bc
\btr{lcc}
\hline
  Parameter       &  $\nu$ Octantis A               & Reference \\ 
\hline
Spectral type     & K0~III                         & (3) \\ 
$V$ (mag)         &  $3.743\pm0.015$              & (1) \\ 
$\mbv$ (mag)      &  $+2.02\pm0.02$               & (2) \\ 
$(B-V)$           &  $0.992\pm0.004$              & (1) \\ 
$H_{\rm p}$ (\hip mag)  &  $3.8981\pm0.0004$      & (3) \\ 
parallax (mas)    &  $45.25\pm0.25$               & (5) \\ 
Mass (\msun)      &  1.61                         & (6) \\ 
Radius (\rsun)    &  $5.81\pm0.12$                & (6) \\ 
\teff (K)         &  $4\,860\pm40$                & (6) \\ 
Luminosity (\lsun)    &  $17.0\pm0.4$             & (2) \\ 
$\log g$ ($\rm g\,cm^{-2}$) & $\rm 3.12\ (\pm0.10\ dex)$ &  (6) \\ 
$\rm [Fe/H] (dex)$&  $+0.18\pm0.04$               & (6) \\ 
$v\sin i$ (\kms)  &  2.0                          & (4),(6) \\ 
Age (Gyr)         &  $\sim2.5$--3                 & (6) \\         
\hline							                     
\etr
\ca{Stellar and orbital parameters for \nuo. 1:\,Mermilliod (1991), 
2:\,present work, 3:\,ESA (1997), 4:\,Costa \oo (2002), 
5:\,Ramm \oo (2009), 6:\,Fuhrmann \& Chini (2012).  
The changes to several stellar parameters given in \drb by Fuhrmann \& Chini 
for the most part originate in their increased estimate for the metallicity 
(shifting Eggen's (1993) value from $\rm Fe/H=-0.11$ to their $+0.18$. Their 
mass errors are claimed to be likely less than 10 per cent. Their 2\si errors 
are halved here to be consistent with the 1\si errors used elsewhere in this 
study.}
\lab{stellar}
\ec
\ete

Besides the possibility that \nuo may be revealing a new type of 
yet-to-be-determined RV-creating surface process, the remaining alternate 
explanation that \nuo instead harbours this particular planet is unexpected 
for at least three reasons: \nuo is substantially tighter than any 
other planet-harbouring binary system ($\abin\sim2.6$~AU; see 
\tl{orbital}), the planet orbit supposedly lies about midway between the two 
stars ($\apl/\abin\sim0.5$), and as a consequence, from orbit stability 
considerations, the Jupiter-mass circumstellar (S-type) orbit must be 
retrograde with respect to the binary orbit (Eberle \& Cuntz 2010) where it is 
more likely to have long-term stability. It is the prediction that the 
planet orbit must be 
retrograde with respect to the stellar orbits that makes the \nuo planet 
so far unique. This geometry is 
fundamentally different from the growing list of transiting exoplanets 
which demonstrate, via the Rossiter-McLaughlin effect, that their orbits can 
also be retrograde, since the R-M effect can only determine anti-alignment 
of a planet orbit's axis with respect to the host star's rotation axis 
(the first such exoplanet being HAT-P-7b: Narita \oo 2009; Winn \oo 2009). 
Other planets may also have retrograde orbits but their true nature is so far 
hidden by the circumstances of their discoveries, such as the absence of an 
observable R-M effect, or the absence of the demands of stability modelling 
as applies to \nuo. Rather than `simple' planet migration to explain all orbit 
evolution scenarios, both types of retrograde geometry imply substantial 
dynamical interactions can have a leading role.

The controversial geometry of \nuo has been investigated with 
increasing depth 
beginning with Eberle \& Cuntz (2010) and subsequently by Quarles, Cuntz \& 
Musielak (2012) and Go{\'z}dziewski \oo (2013). Go{\'z}dziewski \oo explored 
the system in considerable detail and found that stable solutions 
consistent with the Ramm \oo results existed but were confined to tiny 
regions of the phase space. Not only is this 
geometry unprecedented but the formation of such a system is generally 
considered incompatible with theoretical expectations (\eg Paardekooper, 
Th{\'e}bault \& Mellema 2008; Kley 2010; Th{\'e}bault 2011; Rafikov \& Silsbee 
2015). However, the work of, for example Trilling \oo (2007), suggests 
otherwise. They studied infra-red excesses of close main-sequence A3-F8 
binaries -- encompassing the likely spectral range of \nuo's progenitor -- and 
found a large fraction ($>60$ per cent) with small separations ($\abin<3$~AU, 
as has \nuo) had infra-red excesses: about 50 per cent consistent with 
circumbinary debris discs and about 30 per cent 
consistent with circumstellar discs. Occasionally, the deduced debris disks 
were located in apparently unstable orbits. These results they claim suggests 
planet formation in such close binaries may not be so unlikely after all. Thus 
\nuo may well be the first-discovered circumstellar consequence 
of such a precursor system.

\bte
\bc
\btr{lcc}
\hline
Parameter         & Binary   & conjectured planet  \\   
\hline 							                     
${\mc M}\sin i$   &  $0.55$ (\msun)      &  $2.4$ (\mjup)     \\ 
$a$  (a.u.)       &  $2.6\pm0.1$         &  $1.2\pm0.1$       \\  
$K$ (\kms)        & $7.032\pm0.003$      & $0.052\pm0.002$    \\ 
$P$ (days)        &  $1050.1\pm0.1$      &  $417.4\pm3.8$     \\ 
$e$               &  $0.2359\pm0.0003$   &  $0.12\pm0.04$     \\ 
$i$ (\dg)         &  $70.8\pm 0.9$       &  ?                 \\ 
$\omega$ (\dg)    & $75.05\pm0.05$       &  $260\pm21$        \\ 
$\Omega$ (\dg)    &  $87\pm 1.2$         &  ?                 \\ 
$N$ (\# RV observations)  &  \ml{2}{c}{222}                    \\
RMS (\ms)         &  \ml{2}{c}{19}                      \\
\hline
\etr
\ca{Orbital parameters for \nuo from a keplerian fit. All values from \drb, 
except the secondary's scaled mass (Fuhrmann \& Chini 2012).}
\lab{orbital}
\ec
\ete

At the time of writing, approximately 1500 exoplanets have been 
confirmed.\fn{See \eg The Extrasolar Planets Encyclopeadia at 
http://exoplanet.eu/ and Exoplanet Orbit Database at http://exoplanets.org}
Before the upsurge of discoveries from transit-detection programs such as 
HATNet (\eg Bakos \oo 2007), SuperWASP (\eg Collier Cameron \oo 2007), and the 
$Kepler$ mission (see \eg Batalha \oo 2011; Rowe \oo 2014), most were 
discovered by the radial-velocity technique. Now the RV-detection fraction is 
closer to about 30 per cent of the total. Of the 120 or so evolved 
planet-hosting stars, two-thirds are giants and one-third subgiants, the 
first confirmed such examples being $\iota$~Draconis (K2~III: Frink \oo 2002), 
HD\,47536 (K1~III: Setiawan \oo 2003), and the previously enigmatic 
$\gamma$~Cephei~A (K1~IV: Hatzes \oo 2003). Jofr{\'e} \oo (2015) and Reffert 
\oo (2015) describe two recent analyses of large 
samples of evolved stars with and without planets. Approximately 10--15 per 
cent of claims, including \nuo b, are unconfirmed, retracted or controversial.

The number of planets found in multiple stellar systems also continues to 
grow but discoveries there are less frequent, and Doppler-spectroscopic planet 
searches 
are understandably biased against binaries as tightly bound as \nuo. A recent 
review of planets in such systems by Roell \oo (2012) deduced the 
fraction of sytems with exoplanets was then about 12\% (of 477 host 
systems, 47 were binaries and 10 triple). Varied efforts to discover unknown 
stellar companions have routinely identified new components (early 
examples being Patience \oo 2002, Mugrauer \oo 2004, and Raghavan \oo 2006), 
so this fraction cannot be considered likely to be definitive. The majority of 
planets found in binary systems are circumstellar but a few
have been identified in circumbinary (P-type) geometries \eg HW~Vir (Lee \oo 
2009). More recent studies, such as SPOTS (Search for Planets Orbiting Two 
Stars), 
are specifically targeting close binary systems (Thalmann \oo 2014). SPOTS is 
using direct imaging to search for planets in circumbinary orbits in 
small-separation systems comparable to that of \nuo ($\abin\lesssim5-10$~AU).

All of the so-far-discovered circumstellar planets in multiple systems have 
binary separations exceeding that of \nuo by an order of 
magnitude or so, and generally a lot more. The systems that are tightest 
($\abin\sim20$~AU) include $\gamma$ Cephei A (Cambell, Walker \& Yang 1988; 
Hatzes \oo 2003), GJ~86~A (Queloz \oo 2000), HD\,41004~A (Zucker \oo 2004), 
and HD\,196885~A (Correia \oo 2008; Th{\'e}bault 2011). Interestingly, 
$\gamma$~Cep~A (which has stellar properties very similar to \nuo - see 
Fuhrmann 2004), was briefly considered one of the strongest candidates for 
hosting the first 
RV-detected exoplanet until later work regrettably questioned that 
possibility (Walker \oo 1992), and GJ~86~Ab was one of the other earliest 
planet discoveries. Most recently 
$\alpha$~Cen~B ($\abin\sim17.5$~AU) has been claimed to host a very close 
planet (Dumusque 
\oo 2012; $\ppl\sim3.2$~days) but its existence has been challenged 
(Hatzes 2013). Other planets claimed in tight binaries, such 
as that for HD\,188753 ($\abin\sim13$~AU) were later shown to be 
non-existent (Eggenberger \oo 2007). None of these systems have separation 
ratios anywhere near as large as that claimed for \nuo, with $\gamma$~Cep, 
HD\,196885 and HD\,41004 all having $\apl/\abin\sim0.1$.

Thus, \nuo is the seat of one of the many unconfirmed and, in view of its 
geometry, perhaps most extraordinary of the controversial planets. Hence, it 
is appropriate that all available 
investigations be pursued to eliminate or support alternative explanations. 
Whilst \drb provided sound reasons to discredit spots and pulsation as the 
likely cause of the supposed planet signal, besides their bisector analysis 
which was inconsistent with either stellar cause, much of that reasoning was 
qualitative. Our effort here is to provide further quantitative evidence 
for the lack of support for these surface-dynamical causes.

Stellar activity in the form of spots or pulsation are expected to reveal 
themselves in many ways, including radial velocity, photometric, spectral-line 
and surface-temperature variations. The photometric stability of \nuo has 
already been 
mentioned (ESA 1997) and recorded in \tl{stellar}. High-resolution spectra 
obtained for radial velocity purposes 
are ideal for studies of spectral-line profiles, symmetries and depths, 
when, as is the case for \nuo, the spectra are uncomplicated (it is strictly 
SB1) and sharp-lined (due to its late spectral type and low rotation; 
$v\sin i=2$\kms: Costa \oo 2002; Fuhrmann \& Chini 2012).

The most common way spectral-line variations are studied for assessing the 
reality of exoplanets is of the bisectors of the lines themselves (\eg  
Gray 1982, 1983; Gray \& Hatzes 1997) or of the cross-correlation function 
(\eg Queloz \oo 2001). But measurements derived from bisection, as valuable as 
it is, is essentially limited to assessing the profile for asymmetry 
against which velocities can be compared \eg convective velocities Gray (1982)
or suspected planet signals \eg Queloz \oo (2001). The bisector spans can also 
be related in a fairly general way to 
spectral types (Gray 1982; Gray 1992) but there is no apparent opportunity to 
convert these bisector spans quantitatively to a fundamental property such as 
effective temperature.

A star's effective temperature, \teff, is a critical parameter as it has 
fundamental and far-reaching consequences for many other stellar 
characteristics. As a consequence, many methods have been devised to measure 
it. In their comparison of two methods, one based on iron excitation and 
ionization balance and the other the infrared flux method, Tsantaki \oo 
(2013) list several others that include interferometry (using the 
relationship between diameter and luminosity), photometrically derived colour 
indices (which have different dependencies on temperature), and line 
properties such as the H$\alpha$ wings, spectral synthesis strategies 
and, the method we will employ, line-depth ratios (LDRs). Gray (1992) 
also discusses various methods, including LDRs. Whilst LDRs perhaps 
require more reduction effort than some of these methods, they have many 
advantageous qualities. For 
instance, line depths are measured downward from the continuum so they should 
not suffer from zero-point errors, they should be applicable to composite 
spectra, and, unlike photometric methods, are independent of interstellar 
extinction and, within sensible limits, sky quality (Gray \& Brown 2001). 

Over the past couple of decades or so, many papers have been published 
describing the extraordinary sensitivity of ratios of the depths 
of suitably chosen pairs of spectral lines for a star's effective temperature. 
LDRs may only provide accuracy in the tens of degrees, but their sensitivity 
allows precision an order-of-magnitude smaller as each of the following 
references (and many others) will testify. Thus they provide an ideal strategy 
for assessing variability of \teff, and as a result of the corresponding 
implications for monitoring surface dynamics, a worthy tool partnered with 
bisectors for evaluating claimed discoveries of substellar companions.

LDRs 
have been used to study inactive main-sequence (\eg Gray \& Johanson 1991;  
Gray 1994; Kovtyukh \oo 2003), giant (\eg  Gray \& Brown 2001; Kovtyukh \oo 
2006), and supergiant stars (Kovtyukh 2007; Pugh \& Gray 2013) as well as 
active stars (Gray \& Baliunas 1995; Padgett 1996; Catalano \oo 2002) 
including the Sun's 11-year cycle (Gray \& Livingston 1997). In turn, 
these temperatures can be used to monitor and assess starspots (\eg 
active stars; Catalano \oo 2002; O'Neal 2006; Biazzo \oo 2007), 
pulsation cycles (\eg of Cephieds: Kovtyukh \& Gorlova 2000, 
and Antares~A: Pugh \& Gray 2013), and hence to help support claims for the 
non-existence or existence of exoplanets (\eg 51~Peg; Gray 1997; Hatzes, 
Cochran \& Bakker 1998). These latter examples demonstrate the 
relatively long though infrequent history LDRs have had with exoplanet 
research.

Our paper continues to Sect.~2 where the observational details are given, 
together with our reductions, choice of spectra lines and our set of LDRs. 
In Sect.~3 we describe the LDR construction, error management of them, and the 
behaviour of our LDRs with regards to \nuo. In Sect.~4 the temperature 
calibration is described, and in Sect.~5 we discuss the consequences of our 
results that will further challenge the possibility that \nuo's 
RV-perturbation might be caused by conventional spots or pulsation.

\se{Observations and data reduction}
The {\'e}chelle spectra were obtained between 2001 and 2007 at Mt 
John University Observatory (MJUO), New Zealand using the 1-m 
McLellan telescope and \hc, a fibre-fed, vacuum-housed spectrograph 
($P\approx 0.01$~atm; Hearnshaw \oo 2002). The spectrograph is located in a 
thermally isolated and insulated room. Both optical fibres used for the 
spectra described here have core diameters of \mbox{$100~\mu$m}, one with a 
\mbox{$50~\mu$m} microslit on its exit face, which provided resolving powers 
respectively of $R\sim$~41,000 and 70,000. The detector was a \kone-pixel CCD 
with 24-$\mu$m 
pixels, which, to achieve complete spectral coverage, required four separate 
CCD positions. Fortunately, the position chosen to address the compromises for 
maximizing radial-velocity (RV) precision (the original purpose for the 
observations - the 
trade-off between spectral-line density, continuum flux and the CCD's 
efficiency), also recorded a useful fraction of the spectral lines 
often used for line-depth ratio analyses, allowing this subsequent 
research to be undertaken. This CCD position recorded wavelengths 
$\ld\ld \sim 4500$--7200~\ang and approximately 44 orders, $n=81-124$.

The spectra were reduced using the Hercules Reduction Software Package 
({\small HRSP} 
v.2.4; Skuljan 2004) which incorporates standard procedures, including 
background subtraction and cosmic ray filtering, normalization using 
quartz-lamp flat-field spectra following careful continuum-level definition, 
and wavelength calibration using a Th-Ar lamp. The flux weighted mid-time of the stellar exposures were determined using 
an exposure meter. The corresponding dispersion solution for each observation 
was determined using Th-Ar spectra obtained immediately before and after the 
stellar spectrum. The 
complete reduction created one-dimensional spectra having a wavelength range 
in the red (where our line-ratio lines are located) of about 50\ang.

\su{Stellar spectra}
\suu{$\nu$ Octantis}
Many of our \nuo spectra were previously introduced in Ramm (2004), 
when the conjectured \nuo planet was first mentioned, with an extended set 
provided in \drb when the 
first detailed study was reported. Since then, it has been decided that 
18 of the 2009 paper's spectra are of dubious quality for precise RVs and, 
making no significant difference to the results to be now described, 
are here rejected. The reasons for these rejections were based on careful 
inspection of the author's observing logbook which identified several nights 
with suspiciously undesirable observing conditions including very poor 
atmospheric seeing (worse than about $7\pp$), Th-Ar lamp malfunctioning and 
failing that night, poorly timed Th-Ar 
calibration spectra, and/or significant observatory-control malfunctions (such 
as poor dome tracking). Also, significantly, the RVs of 
some other target stars on some of these nights also had atypical, non-random 
behaviour. It has also been realised that an additional 21 spectra had 
been acquired in 2007 but overlooked for the 2009 paper and here 
included.\fn{The additional spectra, archived inappropriately, had been 
acquired over 4 consecutive nights (Feb-Mar 2007) during an unexpected 
exchange of CCDs when a new \kfour~CCD was briefly de-commissioned and the 
\kone~detector returned to service. A significant revision of the 
\kone~spectra's RVs, acknowledging these rejections and additions, combined 
with 
a more recent large RV dataset using an iodine cell (whose lines unfortunately 
contaminate our LDR wavelengths) and their ongoing analysis will be the 
subject of a companion paper in 
preparation.\lab{foot}} Hence, a total of 225 \nuo spectra are analysed and 
discussed in this paper (215 with $R\sim70,\!000$, ten with $R\sim41,\!000$ - 
the 
latter purely for the brief purpose of comparison of the LDRs with resolving 
power).

\suu{LDR-to-temperature calibration stars}
\lab{efornax}
The utility of LDRs is their ability to provide a temperature scale of 
extraordinary precision. Ideally, to calibrate the temperature scale 
as large a set as possible of spectra acquired with the same instrumentation 
and identically reduced is required. Given the significant influence of 
stellar evolution on LDRs (as the cited papers in the Introduction indicate), 
the calibration stars must have a similar evolutionary status ranging over an 
adequate temperature range with the target star's \teff  
somewhere midway within that range so that reliable interpolation of 
temperature, rather than less reliable extrapolation, is possible. We can 
estimate the expected temperature variation for \nuo based on its 
\hip photometry and effective temperature \teff given in \tl{stellar}. 
Assuming the primary star is not pulsating, the Stefan-Boltzmann law predicts 
a corresponding temperature variation of about $\Delta T=10$~K. Thus, a set 
of calibration stars that have temperatures ranging over 
$4860\pm500$~K or so will be adequate (\ie exceeding $\Delta T$ by 2 orders 
of magnitude).

Once again fortuitously, an adequate set of such 
spectra were discovered from past observations with 
\hc and the \kone~detector. These had also been observed during 2001--2007 
with the \kone-CCD and all at $R\sim70,\!000$, and had been acquired for the 
various purposes of RV templates for SB2-spectra and RV-zero-point and RV 
standard-star analyses (Ramm 2004). Hence, whilst sometimes a given star had 
only one spectrum available, it was usually of moderate--high signal-to-noise 
($S/N$). Some properties of the 20 stars 
identified for this calibration task, together with \nuo, are provided in 
\tl{calstars}. The $V$ magnitudes and ($B-V$) colour indices are from 
Mermilliod (1991).  The parallaxes used to derive the absolute 
magnitudes, \mbv, are from van Leeuwen (2007) except for $\beta$~Ret and \nuo 
whose parallaxes were determined with higher precision in \drb\ -- that for 
\nuo is nearly $8\times$ more precise. The photometric variability, 
$\si_{_{\rm HIP}}$, is taken from  
\hip observations (ESA 1997). This was reviewed as a guide to identifying 
potentially unsuitable stars where only a small number of spectra were 
available but which may have had undesirable variability that was not 
adequately sampled with our observations. Several stars are, like \nuo, SB1s. 
These include the star with the largest $\si_{_{\rm HIP}}=0.0029$, HD\,219834, 
for which 18 spectra were chosen that had been acquired over 
646~days.\fn{HD\,219834 has an orbital period $P=6.3$~yrs. Its LDR behaviour 
at all levels of our analysis is consistent with the other calibration stars 
chosen, hence justifying its inclusion. The other SB1s were HD\,23817 
($P=5.3$~yrs), HD\,28307 ($P=16.4$~yrs), HD\,49393 ($P=4.8$~yrs), and 
HD\,194215 ($1$~yr) (Pourbaix \oo 2004). Based on the disparity of published 
RVs and the author's two high-precision RVs, HD\,18907 is also suspected 
of being an SB1 (Ramm 2004).\lab{sbpred}} ESA (1997) is also the source of the 
spectral types (G3--K2) and luminosity classes. The luminosity class of \nuo 
is consistently given as III in Houk \& Cowley (1975), ESA (1997), and Gray 
\oo (2006) though our absolute magnitude and temperature suggest it is less 
evolved and nearer III/IV.

\begin{table*}
\begin{center}
\setlength{\tabcolsep}{2pt}
\btr{rc@{\es}r@{\es}r@{\dgap}c@{\es}c@{\dgap}c@{\dgap}c@{\dgap}c@{\es}r@{\dgap}cc}
\hline
\ml{1}{c}{HD} & \ml{1}{c}{Name} & \ml{1}{c}{$V$}   & \ml{1}{c}{\mbv}  & \mdiff            & $(B-V)_{_0}$     & \tbv         & Spec.     & $\si_{_{\rm HIP}}$ & \ml{1}{c}{\#} & $S/N$ \\
              &                 & \ml{1}{c}{(mag)} & \ml{1}{c}{(mag)} & \ml{1}{c}{(mag)}  & (mag)            &  (K)         &           &   (Hmag)           &                &        \\
\hline
   4128       &   $\beta$ Cet   &  \es2.04         & \es$-0.32$       &  $-7.14\pm0.05$   &  $1.030\pm0.009$ &  $4791\pm19$ & K0~III     &  0.0008            & \dgap 6     & $203\pm34$  \\
  18907       &  $\epsilon$ For &    5.88          &   $  3.34$       &  $-2.29\pm0.05$   &  $0.798\pm0.008$ &  $5328\pm20$ & G8/K0~V(?) &  0.0006            &       1     & \ml{1}{c}{$197$}  \\
  23817       &   $\beta$ Ret   &    3.84          &   $  1.38$       &  $-5.90\pm0.03$   &  $1.141\pm0.009$ &  $4575\pm16$ & K0~IV      &  0.0005            &       4     & $196\pm17$  \\
  25723       &    $-$          &    5.62          &   $  0.26$       &  $-6.84\pm0.22$   &  $1.098\pm0.030$ &  $4657\pm58$ & K1~III     &  0.0008            &       1     & \ml{1}{c}{$199$}  \\
  28307       &  $\theta_1$ Tau &    3.84          &   $  0.46$       &  $-6.11\pm0.09$   &  $0.963\pm0.012$ &  $4931\pm26$ & G7~III     &  0.0006            &      33     & $200\pm42$  \\
  35369       &   29    Ori     &    4.13          &   $  0.71$       &  $-5.99\pm0.09$   &  $0.972\pm0.012$ &  $4913\pm25$ & G8~III     &  0.0004            &       1     & \ml{1}{c}{$473$}  \\
  39364       &   $\delta$ Lep  &    3.78          &   $  1.07$       &  $-5.58\pm0.05$   &  $1.003\pm0.013$ &  $4846\pm27$ & G8~III/IV  &  0.0004            &       2     & $434\pm42$  \\
  49293       &   18    Mon     &    4.47          &   $ -0.80$       &  $-8.00\pm0.23$   &  $1.141\pm0.028$ &  $4574\pm52$ & K0~III     &  0.0006            &       1     & \ml{1}{c}{$496$}  \\
  61935       &   $\alpha$ Mon  &    3.93          &   $  0.65$       &  $-6.13\pm0.08$   &  $1.036\pm0.013$ &  $4779\pm25$ & K0~III     &  0.0004            &       1     & \ml{1}{c}{$500$}  \\
  80170       &      $-$        &    5.32          &   $  0.19$       &  $-7.37\pm0.12$   &  $1.193\pm0.026$ &  $4477\pm49$ & K2~III     &  0.0004            &      48     & $203\pm16$  \\
 100407       &   $\xi$  Hya    &    3.54          &   $  0.54$       &  $-5.97\pm0.07$   &  $0.957\pm0.015$ &  $4946\pm32$ & G8~III     &  0.0006            &       1     & \ml{1}{c}{$379$}  \\
 109492       &      $-$        &    6.22          &   $  2.85$       &  $-2.56\pm0.06$   &  $0.742\pm0.012$ &  $5477\pm33$ & G4~IV      &  0.0006            &       1     & \ml{1}{c}{$184$}  \\
 188376       &   $\omega$ Sgr  &    4.70          &   $  2.63$       &  $-2.74\pm0.05$   &  $0.763\pm0.007$ &  $5420\pm20$ & G3/G5~III  &  0.0007            &       2     & $269\pm92$  \\
 194215       &     $-$         &    5.84          &   $ -0.09$       &  $-7.42\pm0.31$   &  $1.137\pm0.037$ &  $4581\pm70$ & G8~II/III  &  0.0007            &       1     & \ml{1}{c}{$229$}  \\
 203638       &   33    Cap     &    5.36          &   $  1.03$       &  $-6.39\pm0.14$   &  $1.178\pm0.018$ &  $4505\pm33$ & K0~III     &  0.0006            &       8     & $193\pm 5$        \\
\bf{205478}   & \bf{$\nu$ Oct}  & \bf{3.74}        &  $\mathbf{2.02}$ &  $\mathbf{-4.78\pm0.13}$   &  $\mathbf{0.997\pm0.007}$ &  $\mathbf{4858\pm14}$ & \bf{KO~III}     &  \bf{0.0004}            &     \bf{215}     & $\mathbf{204\pm28}$  \\
 219834       &   94    Aqr     &    5.20          &   $  3.57$       &  $-2.14\pm0.24$   &  $0.794\pm0.011$ &  $5338\pm29$ & G6/G8~IV   &  0.0029            &      18     & $172\pm17$  \\
 220957       &    $-$          &    6.38          &   $  0.89$       &  $-5.19\pm0.23$   &  $0.924\pm0.032$ &  $5018\pm71$ & G6/G8~III  &  0.0012            &       1     & \ml{1}{c}{$202$}  \\
 222803       &    $-$          &    6.08          &   $  1.88$       &  $-4.90\pm0.12$   &  $1.000\pm0.018$ &  $4853\pm36$ & G8~IV      &  0.0007            &       1     & \ml{1}{c}{$202$}  \\
 222805       &    $-$          &    6.06          &   $  2.72$       &  $-3.59\pm0.06$   &  $0.921\pm0.011$ &  $5025\pm26$ & G8~IV      &  0.0007            &       1     & \ml{1}{c}{$214$}  \\
 223807       &    $-$          &    5.75          &   $ -0.63$       &  $-8.10\pm0.33$   &  $1.218\pm0.046$ &  $4431\pm86$ & K0~III     &  0.0005            &       1     & \ml{1}{c}{$226$}  \\
\hline															       
\etr															       
\caption{Some parameters and observation statistics for $\nu$ Octantis and 
the calibration stars used for the conversion of the line-depth ratios to 
temperatures. \tbv was derived using \eql{gform}, \# is the number of spectra 
used and $S/N$ is their mean signal-to-noise. Some of our results cast 
suspicion on the correct spectral class of HD\,18907, which appears to be 
evolved beyond class V (see also \fnl{sbpred} and \fnl{spsub}).}
\lab{calstars}
\end{center}
\end{table*}

The colour index ($B-V$) was used to estimate the effective temperatures. 
Several relations of this type were compared in Strassmeier \& Schordan 
(2000) in their study of LDRs of Morgan-Keenan class III giants. Of 
those compared the relation 
chosen here (from Gray 1992) was found to be of comparable accuracy to the 
others for our (B-V) range $\sim0.7-1.2$:

\bea
\lab{gform}
\log \tbv &=& 3.988 - 0.881(B-V)_{_0} + 2.142(B-V)_{_0}^2 \nonumber \\
           & &  - 3.614(B-V)_{_0}^3 + 3.2637(B-V)_{_0}^4   \nonumber  \\ 
           & &  - 1.4727(B-V)_{_0}^5 + 0.2600(B-V)_{_0}^6\ .
\eea
\noindent
The conversion of the $(B-V)$ values from Mermilliod (1991) to the 
intrinsic values $(B-V)_{_0}$ given in \tl{calstars} was determined using an 
extinction correction that does not make allowance for galactic 
latitude. The reasons for this decision include: 1.\,the stars are mostly 
relatively close (mean distance $\sim70\pm50$~pc), 2.\, only two stars have a 
galactic latitude $<5\dg$ (HD\,49293 and HD\,109492), and only the former is 
beyond 100~pc, and 3.\,the simple isotropic formula we used predicted 
HD\,49293's temperature to be consistent with published values (\eg 
Ammons \oo 2006.) The formula is from Henry \oo (2000): 
$E(B-V)=0.8/3.3 = 0.2424~\rm mag\,kpc^{-1}$, where $E(B-V)$ is 
$A_{_{\rm V}}=0.8~\rm mag\,kpc^{-1}$, the interstellar $V$ absorption, divided 
by the ratio of total to selective extinction.

An estimate of the stage of evolution (corresponding to a gravity index) was 
calculated by taking the difference between the 
absolute magnitude and the star's zero-age main-sequence (ZAMS) magnitude at 
that temperature (see \eg Catalano \oo 2002). 
A cubic polynomial fit was made to the ZAMS values given in Allen (1991) and 
this locus was compared to the \mbv vs. \tbv values of nearly 2000 \hip 
stars within 80~pc, again using 
\eql{gform}. The locus follows the \hip distribution so closely that at most 
the only difference an alternative ZAMS locus could imply would be a minor 
zero-point offset which cannot effect our results significantly. The ZAMS 
magnitude is therefore considered to have zero error for the subsequent error 
estimates that will be based on standard error-propagation principles. The  
H-R diagram relating \mbv to \tbv of the LDR-calibration stars and 
\nuo is given in \fl{hrd}. The position of \nuo is seen to be about midway 
between those of the calibration stars as is preferred. Another check of 
the suitability of the LDR-calibration stars is that all are at least two 
magnitudes evolved from the ZAMS. These values for 
\mdiff are given in \tl{calstars}, together with the number of spectra used 
and their mean signal-to-noise $S/N$ in the vicinity of the LDR lines. For the 
20 calibration stars, the average $S/N$ is $270\pm115$. It would 
appear from these preliminary results that the star HD\,18907 is unlikely to 
be properly classed as a dwarf since its high galactic latitude ($g=-61\degr$) 
and 
proximity ($d=32$~pc) makes interstellar reddening an unlikely complication 
to its H-R diagram location.\fn{HD\,18907 also appears better placed as a 
subgiant since, as will be soon become apparent, its LDR behaviour is also 
consistent with it being somewhat evolved.\lab{spsub}}

\bfi
\bc
\rotatebox{-90}{\scalebox{0.3}{\includegraphics{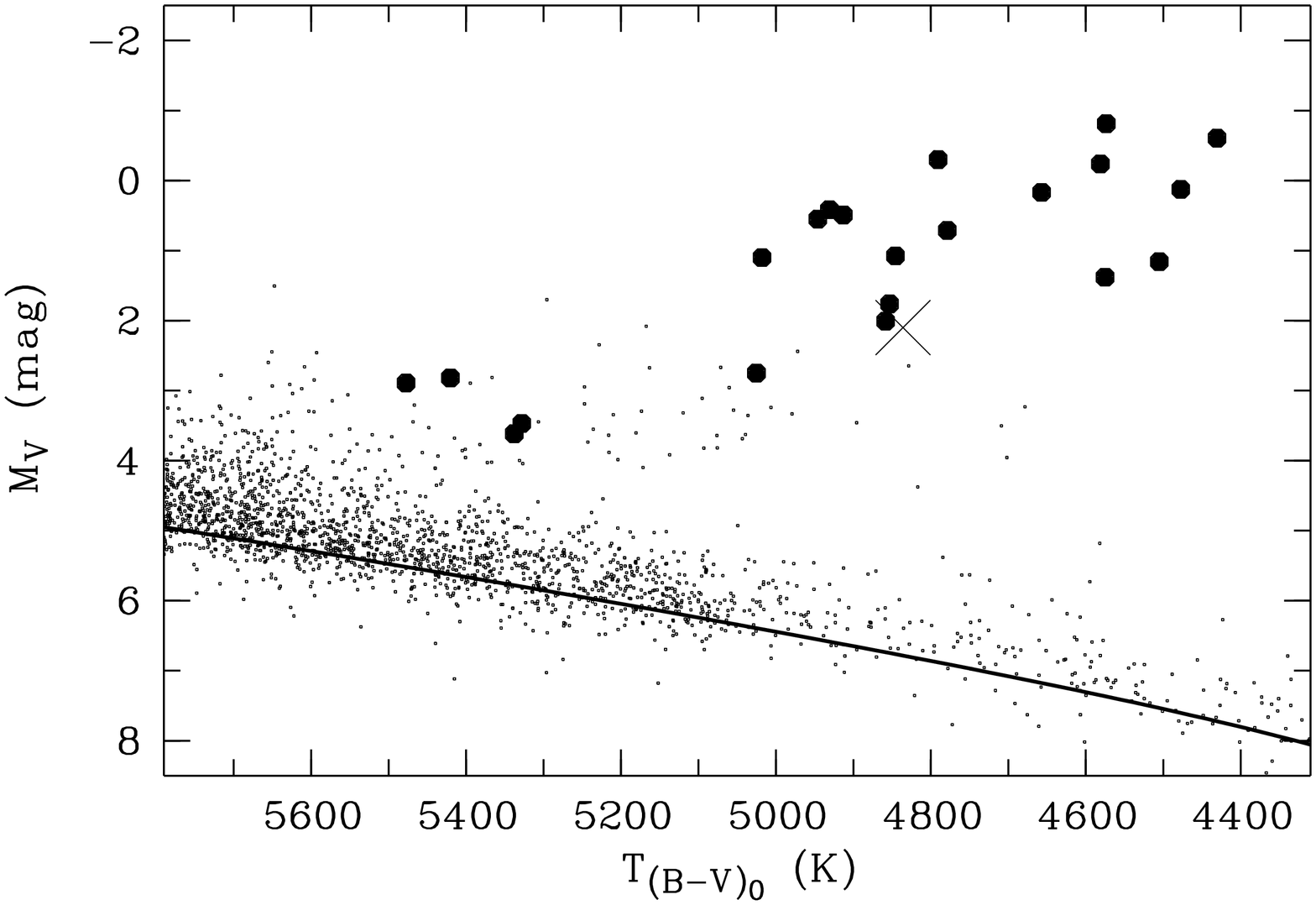}}}
\caption{The absolute magnitudes and \tbv-calibrated temperatures of 
20 stars `$\bullet$' and $\nu$~Octantis identified with the `$\times$'. 
Approximately 2000 \hip stars within 80~pc are included to demonstrate the 
adequacy of our ZAMS locus (solid line).}
\label{hrd}
\ec
\efi

\su{Spectral lines for depth ratios}
\lab{rotation}
Our spectra include two relatively distinct regions that include lines often 
used for LDR analyses. One region has the approximate range 6410--6460\AA\ 
(see \eg Strassmeier \& Schordan 2000) and the other 6200--6275\AA\ (\eg Gray 
\& Johanson 1991; Gray \& Brown 2001; Catalano \oo 2002; Biazzo \oo 2007). 
Other studies examine ratios over much wider wavelength ranges \eg Kovtyukh 
(2007) studied supergiants using bluer lines ranging from 5350--6080\AA. Our 
\kone~detector restricts our choices in the 6410--6460\AA\ range to only four 
lines, all of which also have a relatively high potential, $2.5< \chi <5.6$~eV 
making these a poor choice. The second region 6200--6275\AA, however, includes 
10 lines all recorded in one spectral order, $n=91$, and present on all 
spectra.\fn{Doppler-shifting of the lines for our SB1s could potentially move 
any line close to the order edge off our detector. Attempts to use an 
additional two lines at $\sim6266$~\AA\ were 
thwarted as they were not recorded after the detector exchange described in 
\fnl{foot} and so were discarded since all ratios were wanted for all 
observations.}

Following the recommendations of Kovtyukh \oo (2006) we selected iron-peak 
elements (Si, V, Fe) lines that are less gravity dependent and are expected to 
have less star-to-star variations in element abundances. A line's behaviour 
with temperature differences can be 
qualitatively predicted from its excitation potential, $\chi$. The smaller 
the value, the greater will be the line's growth be with temperature. Four of 
our lines are fast-growth V\sci lines with low $\chi$ 
($\chi_{\rm low}\sim0.3$~eV) and the other six are slower-growth Fe\sci, 
Fe\scii 
and Si\sci lines with $\chi$ much higher ($\chi_{\rm high}$ in the range 
2.4--5.6~eV). The nominal wavelength and excitation 
potential of each line, listed in \tl{expot}, has been obtained from the VALD3 
database whose original sources were Kurucz (2007, 2009, and 2013).

\begin{table}
\bc
\setlength{\tabcolsep}{2pt}
\btr{c@{\dgap}c@{\dgap}c@{\dgap}c}
\hline
$\lambda $ (\AA) & Element & $\chi $ (eV) & label \\
\hline
6232.64  &  Fe\sci   &  3.65  & 6232 \\
6233.20  &  V\sci    &  0.28  & 6233 \\
         &   &  &  \\		                
6242.83  &  V\sci    &  0.26  & 6243 \\
6243.82  &  Si\sci   &  5.62  & 6244 \\
6246.32  &  Fe\sci   &  3.60  & 6246 \\
6247.56  &  Fe\scii  &  3.89  & 6247 \\
         &   &  &  \\		                
6251.83  &  V\sci    &  0.29  & 6252 \\
6252.55  &  Fe\sci   &  2.40  & 6253 \\
         &   &  &  \\		                
6255.76  &  Fe\sci   &  4.45  & 6256 \\
6256.89  &  V\sci    &  0.27  & 6257 \\
\hline
\etr
\caption{Spectral lines used for line-depth ratios. Included is the integer 
label used in this paper.}
\lab{expot}
\ec
\end{table}

We constructed as many ratios as our line list permitted and 
ultimately selected four sets of six ratios that gave reliable correlations 
with \tbv, pairing each low-$\chi$ line with one of the six higher-$\chi$ 
lines, all with $\chi_{\rm low}$ as the numerator. This 
choice ensured the $T$-LDR plots for our calibrations stars were always 
approximately linear. Inverting the ratio results in exponentially varying 
distributions, a more complex task for fitting regression curves. \fl{spectra} 
illustrates the location of the lines and their behaviour with 
effective temperature for \nuo and two calibration stars, the IAU RV-standard 
star HD\,80170 (Udry, Mayor \& Queloz 1999), and HD\,188376. As can be seen, 
and is the case for all the 
stars studied here, the lines are very sharp indicating very low $v\sin i$ 
similar to \nuo. Thus we do not expect any rotational broadening complications 
to our ratio measurements (see \eg Biazzo \oo 2007). The complete list of 
ratios is given in \tl{ratiostats}, and as the paper progresses, the 
accompanying columns will be described.

\bfi
\hspace{-3em}
\rotatebox{-90}{\scalebox{0.36}{\includegraphics{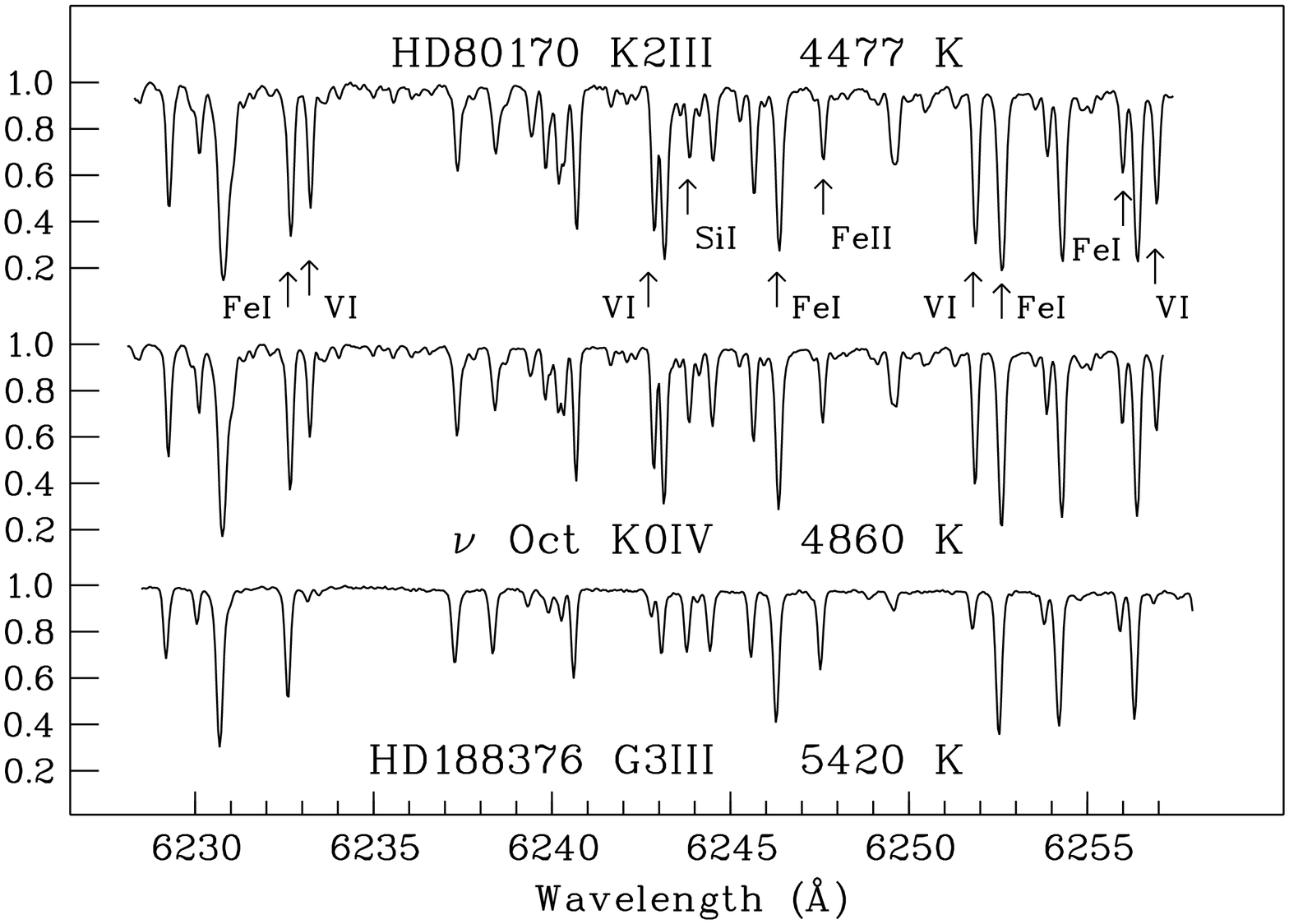}}}
\caption{Three sample spectra illustrating the variations of our ratio lines 
with effective temperature and the resolution of them.}
\label{spectra}
\efi

\begin{table*}
\begin{center}
\setlength{\tabcolsep}{2pt}
\btr{l@{\dgap}c@{\es}r@{\dgap}r@{\dgap}@{\dgap}c@{\dgap}c@{\dgap}r@{\dgap}@{\dgap}c@{\dgap}@{\dgap}l}
\hline
  Ratio &   $r\ldr\pm\vep_r$ & \ml{1}{c}{$s\ldr$}              & \ml{1}{c}{$\delt\ldr$}& \ml{1}{l}{$\nabla T\ldr$} & $\si\ldr$    &  \ml{1}{c}{$s\mldr$}          & \ml{1}{l}{$\nabla T\mldr$}    & $\si\mldr$  \\
        &                    & \ml{1}{c}{($\times0.01$~K)}    &  \ml{1}{c}{(K)}       & \ml{1}{l}{(K)}            &  (K)         &  \ml{1}{c}{($\times0.01$~K)} &      \ml{1}{l}{(K)}           & (K)    \\
\hline
  3332  &   $0.626\pm0.007$  & \es $ -11.0$    &\tg   8   &  $ -123$  &  108   &\dgap$  -9.1$     &  $ -59$ &   57    \\
  3344  &   $1.184\pm0.022$  &     $  -5.4$    &     12   &  $  -85$  &  111   &     $  -4.4$     &  $ -28$ &   55    \\
  3346  &   $0.561\pm0.007$  &     $ -11.7$    &      8   &  $ -121$  &  109   &     $  -9.7$     &  $ -57$ &   59    \\
  3347  &   $1.226\pm0.021$  &     $  -4.8$    &     10   &  $ -115$  &  143   &     $  -3.7$     &  $ -31$ &   55    \\
  3353  &   $0.503\pm0.006$  &     $ -13.2$    &      8   &  $ -131$  &  112   &     $ -10.9$     &  $ -64$ &   61    \\
  3356  &   $1.183\pm0.022$  &     $  -6.9$    &     15   &  $  -91$  &  147   &     $  -4.9$     &  $  -5$ &   76    \\
  4332  &   $0.852\pm0.007$  &     $ -11.3$    &      8   &  $ -146$  &   93   &     $  -9.0$     &  $ -79$ &   53    \\
  4344  &   $1.611\pm0.025$  &     $  -5.2$    &     13   &  $  -90$  &  112   &     $  -4.0$     &  $ -28$ &   57    \\
  4346  &   $0.763\pm0.006$  &     $ -12.1$    &      8   &  $ -145$  &   85   &     $  -9.7$     &  $ -82$ &   47    \\
  4347  &   $1.669\pm0.026$  &     $  -4.6$    &     12   &  $ -130$  &  145   &     $  -3.4$     &  $ -49$ &   60    \\
  4353  &   $0.685\pm0.005$  &     $ -13.7$    &      7   &  $ -159$  &   87   &     $ -11.0$     &  $ -91$ &   49    \\
  4356* &   $1.609\pm0.025$  &     $  -6.0$    &     15   &  $  -68$  &  222   &     $  -3.5$     &  $ +34$ &  113    \\
  5232  &   $0.946\pm0.008$  &     $ -12.3$    &     10   &  $ -151$  &   95   &     $  -9.4$     &  $ -79$ &   57    \\
  5244  &   $1.788\pm0.026$  &     $  -5.4$    &     14   &  $  -85$  &  120   &     $  -4.0$     &  $ -21$ &   65    \\
  5246  &   $0.847\pm0.006$  &     $ -13.0$    &      8   &  $ -150$  &   80   &     $ -10.3$     &  $ -86$ &   48    \\
  5247  &   $1.852\pm0.027$  &     $  -4.7$    &     13   &  $ -133$  &  154   &     $  -3.4$     &  $ -51$ &   67    \\
  5253  &   $0.760\pm0.005$  &     $ -15.0$    &      8   &  $ -168$  &   82   &     $ -11.7$     &  $ -97$ &   49    \\
  5256* &   $1.786\pm0.024$  &     $  -5.7$    &     14   &  $  -33$  &  276   &     $  -3.1$     &  $ +67$ &  130    \\
  5732  &   $0.571\pm0.008$  &     $ -10.1$    &      8   &  $ -114$  &  124   &     $  -8.4$     &  $ -45$ &   64    \\
  5744  &   $1.080\pm0.018$  &     $  -5.1$    &      9   &  $  -79$  &  123   &     $  -4.1$     &  $ -17$ &   58    \\
  5746  &   $0.512\pm0.007$  &     $ -10.8$    &      8   &  $ -112$  &  125   &     $  -9.0$     &  $ -43$ &   66    \\
  5747  &   $1.119\pm0.021$  &     $  -4.6$    &     10   &  $ -104$  &  151   &     $  -3.5$     &  $ -19$ &   58    \\
  5753  &   $0.459\pm0.006$  &     $ -12.1$    &      7   &  $ -120$  &  128   &     $ -10.1$     &  $ -49$ &   67    \\
  5756  &   $1.079\pm0.017$  &     $  -6.5$    &     11   &  $  -91$  &  137   &     $  -4.9$     &  $ -13$ &   66    \\
\hline
\etr															       
\caption{The 24 line-depth ratios, their mean values $r$ for $\nu$ Octantis, 
the slope $s$ of the temperature-calibration-star linear fits, the implied 
sensitivity from those fits $\delt$ for $\Delta r=0.01$, the corresponding 
differences, $\nabla T$, for \nuo of the mean regression temperatures from 
$\tbv=4860$~K, and the standard deviations \si of the calibration-star 
temperature differences from their \tbv values. The two ratios marked with 
`$\star$' were discarded from the final analysis.}
\lab{ratiostats}
\end{center}
\end{table*}

\se{Line-depth ratio measurements}
Each line depth, $D$, is derived based on the local continuum 
level, $S_{\rm c}$, and the line's minimum flux, $S_{\rm p}$, as 
determined by a parabolic fit to the three lowest bins in each line core:

\beq
      D = 1 - \f{\sigp}{\sigc}\ .
\eeq
\noindent
Strassmeier \& Schordan (2000) assessed the relative merits of several 
techniques for this purpose (pixel 
minimum, parabolic fit, gaussian-fit minimum and two equivalent width 
strategies) and concluded in agreement with Gray (1994) that the parabolic-fit 
method was the most internally consistent. Using only the three lowest bins 
assures us that the core can be no less deep than the minimum bin value, which 
is not necessarily the case if more bins are used. We could restrict our 
analysis to lines that are close together (within an \AA\ or so) which may 
minimize errors arising from the continuum estimate, but this would reduce the 
number of usable ratios. Furthermore, as Gray \& Brown (2001) point out, 
amongst the advantages of measuring LDRs is that the line depths are measured 
downward from the continuum so they should suffer very little from 
zero-point errors, particularly if they are measured in a consistent manner.

\begin{figure*}
\hspace{-3em}
\rotatebox{-90}{\scalebox{0.55}{\includegraphics{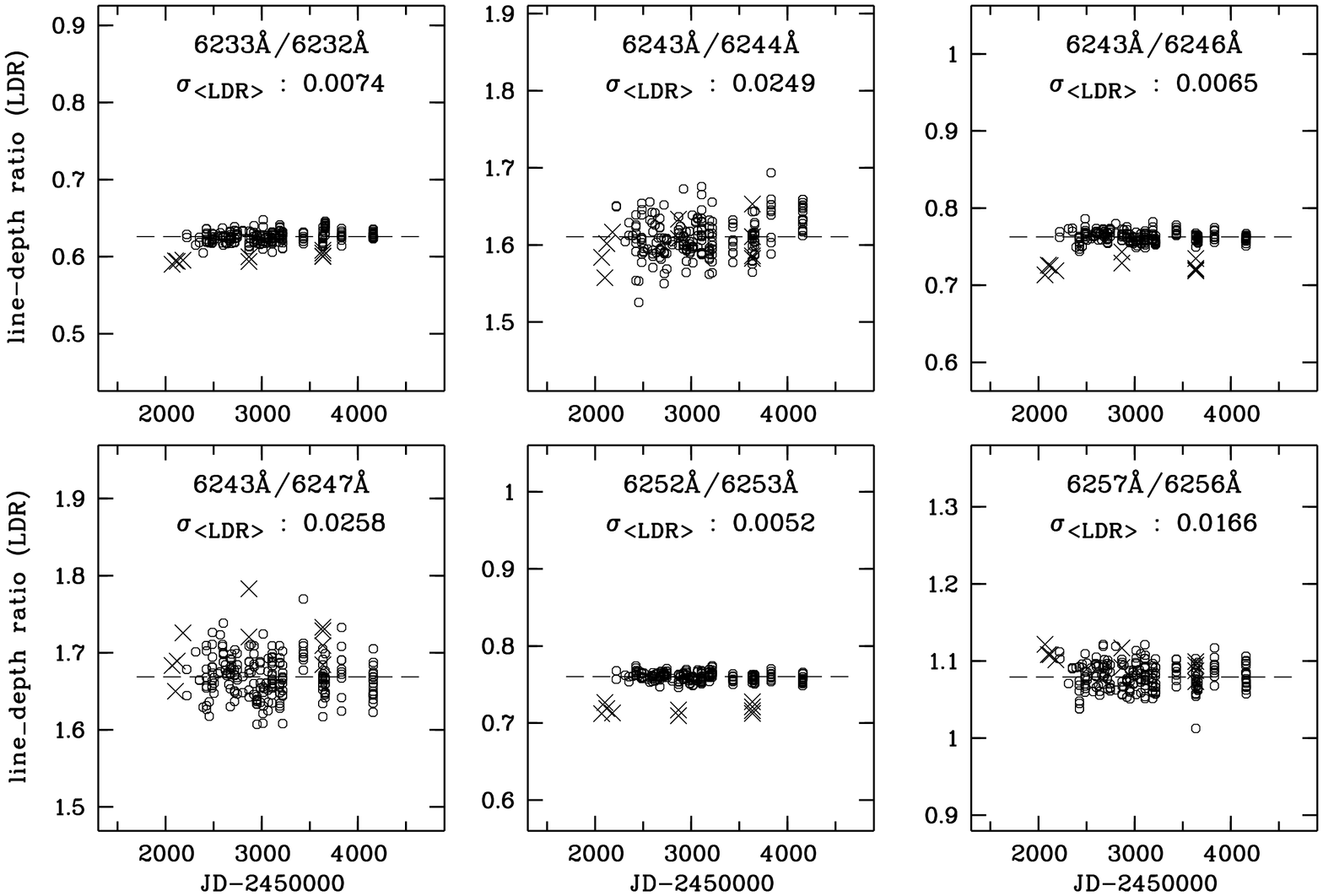}}}
\caption{The time variations of a sample of six line-depth ratios typical for 
$\nu$ Octantis acquired between 2001--2007. The spectra are divided between 
ten acquired at $R\sim41,\!000$ `$\times$', and 215 at $R\sim70,\!000$ 
`$\circ$'. The standard deviation of the weighted mean for the higher 
resolution spectra is included.  The $y$-axis has the same range in each plot.}
\label{nuoldrs}
\end{figure*}

Each of our spectra span about 1000 pixels. We divided this interval into 
100-pixel segments and identified the pixel in each segment with the highest 
value. We fit a low-order polynomial to these peak values and from the 
coefficients determined the continuum level corresponding to each core 
position. Every continuum locus was assessed graphically by eye to confirm 
it was well-behaved.

The errors, \vep, on the depth and ratio, $r$, are calculated as follows:

\beq
      \vep_{_{\rm D}} = \f{\sigp}{\sigc}\sqrt{\f{1}{\sigp} + \f{1}{\sigc}}
\eeq

\beq
      \f{\vep_{\rm r}}{r} = \sqrt{\left(\f{\vep_{_{\rm D_1}}}{D_1}\right)^2 + 
                          \left(\f{\vep_{_{\rm D_2}}}{D_2}\right)^2} 
\eeq
\noindent
where \sigp and \sigc are in ADU (Analogue-to-Digital Unit).

Where more than one spectrum was available for a star, the weighted mean 
ratio $<r>$ for the $N$ spectra was calculated:

\beq
        <r> = \f{\sum_{i=1}^{N}w_ir_i}{\sum_{i=1}^{N}w_i}
\eeq
\noindent
where $w_i=1/\vep_{\rm i}^2$, and assuming the errors are normally 
distributed, the error on the mean is

\beq
        \vep_{<r>} = \sqrt{1/\sum_{i=1}^{N}w_i}
\eeq
\noindent
The final statistical equation used is the standard deviation of the 
weighted mean, which for $k$ ratios is given by

\beq
      \si_{<r>} = \sqrt{\f{\sum_{i=1}^{k}w_i(r_i - <r>)^2 }{\sum_{i=1}^{k}w_i}
                  \f{k}{k-1}}\ .
\eeq

Six typical LDRs and their time variations for \nuo from 2001-2007 are 
presented in \fl{nuoldrs}. There are two significant periods we are searching 
for evidence of in our LDRs: 1.\,that of the $\sim400$~day RV perturbation, 
\ppert , and 2.\,the rotational period of the primary star, \prot, which also 
remains uncertain. Our estimate for the latter is little different from the 
value derived 
by \drb since both Costa \oo (2002) and more recently Fuhrmann \& Chini (2012) 
propose the same 
value for $v\sin i=2\kms$, but neither assign an error to it. Fuhrmann \& 
Chini revised the radius down slightly to $\mc{R}=5.8\pm0.12\rsun$. 
If we assume conservatively that the error on $v\sin i$ is $\pm0.5$\kms we 
derive $\prot\simeq140\pm35$~days, not dissimilar from \drb. To match \nuo's 
perturbation period of 400~days requires $v\sin i\sim 0.7\kms$. If a 
reliable period can be found from our LDR analysis it might at least resolve 
this uncertainty.

As it turns out, Lomb-Scargle periodograms of the 24 LDRs revealed no evidence 
of any particular period consistently having a significant power exceeding the 
noise, and 
certainly not in the vicinity of \ppert ($380<P<420$~days) or \prot (see 
\fl{periods}). Twenty-one ratios have their peak power at a period $<50$~days 
and these have an average of $18\pm11$~days. In the vicinity of \prot 
($105<P<175$~days) there are two ratios whose peak power falls in 
this range ($r4346$ and $r4353$), both having a period of about $P=134$~days. 

\bfi
\rotatebox{-90}{\scalebox{0.3}{\includegraphics{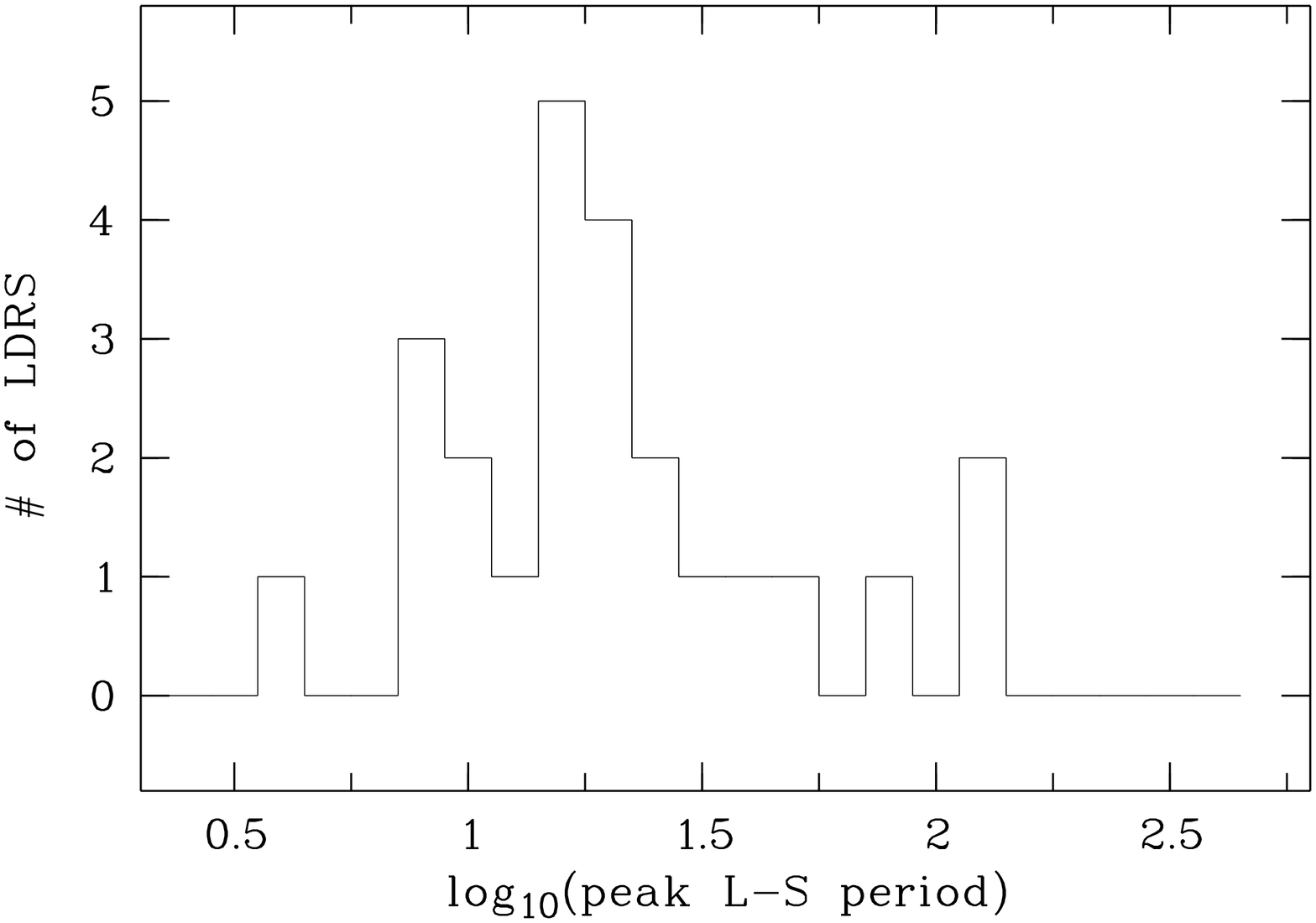}}}
\caption{The distribution of Lomb-Scargle periodogram peak periods for 
24 line-depth ratios. No peak has significant power above the background 
noise.}
\label{periods}
\efi

The results for the six ratios provided in \fl{nuoldrs} have been chosen to 
illustrate three with 
high scatter and three with low scatter. The ratio distributions, $r_{_1}$  
and $r_{_2}$, correspond to the two resolving powers $R$ used. The 
scatter is largely a result of the quotient calculation, so if the 
ratio was inverted, the scatter would re-scale accordingly. Thus, this feature 
has no other particular significance. Another feature is that the difference 
between the means ($<r_{_1}>-<r_{_2}>$) is quite highly correlated with either 
ratio. For instance with $<r_{_2}>$, $\rho=0.85$: the low-scatter $r<1$ ratios 
have negative mean differences and the higher-scatter $r>1$ ratios have 
positive mean differences. With regards our intention to obtain high-precision 
temperatures, one more detail has more relevance: the ratio of the standard 
deviations of those means, $\si_{\rm r_1}/\si_{\rm r_2}$, is approximately 
normally distributed, its mean being $1.0\pm0.3$. Thus, there is apparently no 
significant advantage for the task of getting precise temperatures for \nuo 
with regards these LDRs and these two resolving powers. However, so as to 
avoid apparently 
unnecessary complications for only ten additional observations, as well 
as zero-point offsets that are nevertheless usually present, the lower 
resolving power spectra are no longer included in what follows.

Support for our strategies used to derive the depths, ratios and their 
errors comes from a comparison of these values for our final 215 \nuo spectra: 
the standard deviation of each ratio's mean and the corresponding average 
error are about equal: ($\si_{<r>}/<\vep_{\rm r}>\,\sim0.9$). 

\se{Converting depth ratios to temperatures}
We begin by noting that the behaviour of each line with regards to 
temperature is related to varying degrees to the same stellar properties that 
determine a star's position on an H-R diagram. These properties include its 
absolute magnitude, effective temperature, metallicity, age, mass, 
surface gravity and so on. When a simple approximate dependence (typically 
represented by some low-order polynomial) exists between the LDRs and, say, 
the temperature, the influence of other attributes create scatter about that 
polynomial fit. Since the exact interaction of each of these properties with 
each
other and the LDR still has some uncertainty associated with it, we cannot 
hope to make an accurate allowance for any of them. Consequently, more easily 
and accurately determined parameters are used such as 
the temperature guides $(B-V)$ and $(R-I)$ indices (\eg Flower 1996; 
McWilliam 1990; Gray 1992; Ram\'{i}rez \& Mel\'{e}ndez 2005), whilst for a 
guide to evolution, they include surface gravities, and the parameter we have 
already discussed - and will use - the 
difference between the absolute magnitude and the ZAMS magnitude, \mdiff.

\begin{figure*}
\hspace{-3em}
\rotatebox{-90}{\scalebox{0.55}{\includegraphics{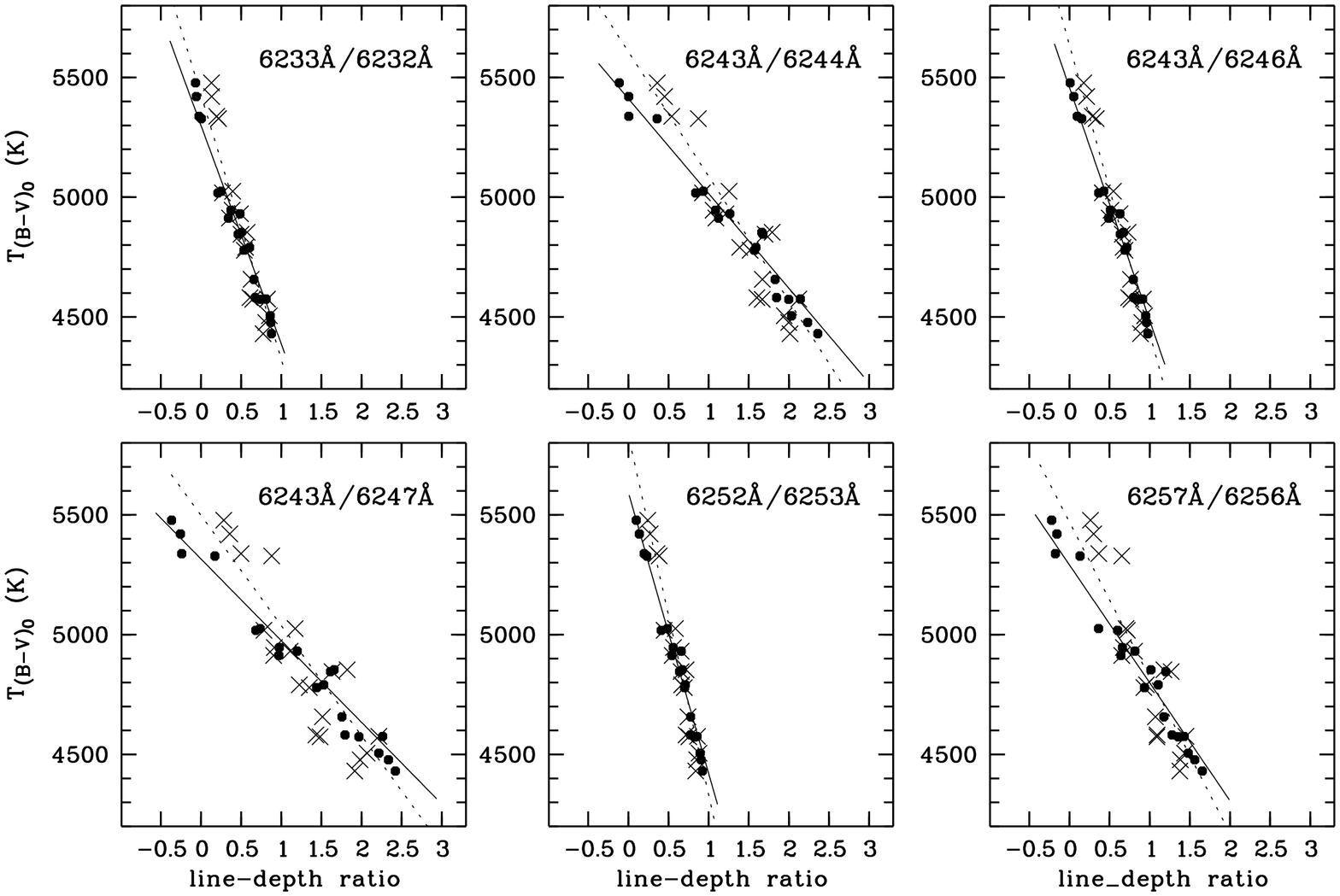}}}
\caption{The variation of six typical line-depth ratios with the 
\tbv-temperatures for 20 calibration stars. The LDRs 
are identified with `$\times$' and a dotted line, whilst 
the `$\bullet$' symbols and solid line identify the modified LDRs (MLDRs) 
subsequently used for deriving the final temperatures for $\nu$~Octantis.}
\label{ldrmldr}
\end{figure*}

The strategy we use for the most part follows that described 
in Catalano \oo (2002) who studied dramatic spottedness of three recognized 
RS CVn-type active binaries. We will derive LDR-calibrated temperatures for 
all our stars, and, as a test of our accuracy, also attempt to recover the 
original values.\fn{It seems inappropriate to claim 
we are actually checking the final `accuracy' of our temperatures, since, even 
a casual review of the literature will often reveal wide variations of 
estimates for \teff for each star, even for the same method of deriving it.} 
O'Neal (2006) has subsequently drawn attention to 
several concerns with regards this method when the stars have higher 
rotation, significant spottedness, cooler temperatures ($\lesssim 4000$~K -- 
due to the contribution of TiO molecular bands in the LDR wavelength range 
employed by Catalano \oo) and inadquate spectral resolving power. We 
believe we will not be adversely affected by these 
important details as all our stars are expected to be inactive, non-spotted, 
slow rotators, and have $\tbv>4400$~K. Our careful inspection of our lines' 
behaviour 
with changing temperature give us confidence we have selected lines for which 
our methods are applicable.

For each ratio, we now plot the mean LDR for all 20 calibration stars with 
respect to their 
temperatures \tbv (see \fl{ldrmldr} for six typical 
examples, the same ratios shown in \fl{nuoldrs} for \nuo). Two fits were 
derived for each ratio's data, one linear (since most ratios had this type of 
distribution) and the second parabolic (since a small fraction of our 
distributions were better fit with this function). After the analysis was 
carried through to its completion, it was evident that 
parabolic fits to the LDR-\tbv distributions actually gave less accurate 
temperature predictions for our calibration stars for all ratios. In fact this 
is not too surprising as LDR papers describing much wider ranges of $(B-V)$ 
(and hence temperature) (\eg Strassmeier \& Schordan 2000; Gray \& Brown 
2001), show essentially linear distributions in the range of our LDR and 
\tbv values. Therefore, 
the analysis that follows always uses linear fits with consideration taken for 
errors in both coordinates (Press \oo 2002).

\su{Temperature predictions from LDRs}
\lab{first}
 each calibration star, the regression-line temperature $T\ldr$ 
corresponding to each ratio was derived, and the mean temperature from all 
ratios ascertained. We can judge the accuracy of the temperature predictions 
from this first 
linear fit by comparing the means, $<T\ldr>$, of our 20 calibration stars to 
their \tbv:

\beq
\nabla T = <T\ldr>\ - \ \tbv\ .
\eeq

\noindent
The mean absolute value is $102\pm56$~K (see \tl{ldrtemps}). This illustrates 
the relative weakness of raw LDRs for temperature accuracy. Our next steps, 
though, improve our accuracy by slightly more than a factor of two.

\begin{table}
\bc
\setlength{\tabcolsep}{2pt}
\btr{c@{\dgap}c@{\dgap}r@{\dgap}r}
\hline
    HD  & \tbv  & \ml{1}{l}{$\nabla T\ldr$} &\ml{1}{c}{$\nabla T\mldr$} \\
\hline
   4128 &   4791   & $   91$         &      $ -25$  \\
  18907 &   5328   & $ -219$         &      $ -77$  \\
  23817 &   4575   & $  -50$         &      $  15$  \\
  25723 &   4657   & $  108$         &      $  46$  \\
  28307 &   4931   & $   19$         &      $ -45$  \\
  35369 &   4913   & $  125$         &      $  42$  \\
  39364 &   4846   & $  -80$         &      $ -68$  \\
  49293 &   4574   & $  190$         &      $  49$  \\
  61935 &   4779   & $   69$         &      $  32$  \\
  80170 &   4477   & $   94$         &      $  52$  \\
 100407 &   4946   & $   96$         &      $  14$  \\
 109492 &   5477   & $  -88$         &      $ -28$  \\
 188376 &   5420   & $  -69$         &      $ -11$  \\
 194215 &   4581   & $  209$         &      $ 100$  \\
 203638 &   4505   & $   31$         &      $  64$  \\
 219834 &   5338   & $  -53$         &      $  67$  \\
 220957 &   5018   & $   79$         &      $  31$  \\
 222803 &   4853   & $ -138$         &      $ -64$  \\
 222805 &   5025   & $  -65$         &      $  39$  \\
 223807 &   4431   & $  159$         &      $  62$  \\
\hline
        &  Mean $\mid\nabla T\mid$ & $102\pm56$ & $46\pm23$  \\        
\hline
\etr
\caption{The differences $\nabla T$ between \tbv and the LDR and MLDR 
regression-line temperatures for the 20 calibration stars.}
\lab{ldrtemps}
\ec
\end{table}

\bfi
\bc
\rotatebox{-90}{\scalebox{0.3}{\includegraphics{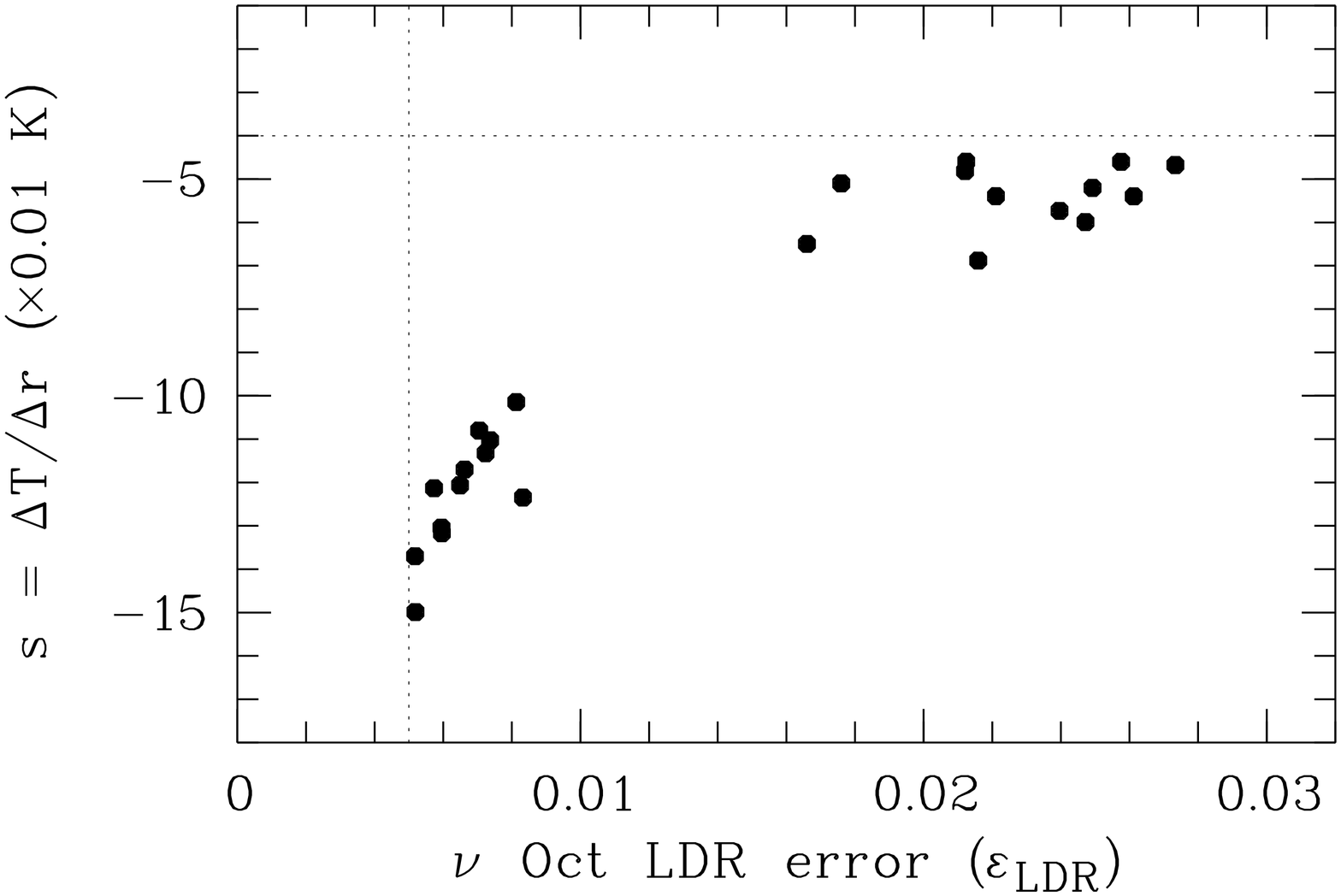}}}
\caption{The relationship between the LDR calibration-star regression-line 
slope, $s$, and \nuo's mean ratio error, \vep. An LDR error bound of about 
0.005 and a slope-temperature bound of about $-4$~K appears to be present.}
\label{sensitivities}
\ec
\efi

We can also judge the temperature sensitivity (and ultimate precision 
potential) of 
each ratio for a small shift \mbox{$\Delta r=0.01$}, by calculating the slope 
of each LDR-\tbv regression line, $s\ldr=\Delt/\Delta r$ and multiplying its 
absolute value by the error on each mean ratio for the 215 \nuo spectra, 
$\vep_{\rm r}$ ($s\ldr$ and $\vep_{\rm r}$ are given in \tl{ratiostats}). This 
detail differs from Catalano \oo (2002) who describe each 
ratio's sensitivity in terms only of the regression-line's slope. However, as 
\fl{sensitivities} illustrates (at least for our stars), there is a relatively 
strong relationship 
between this calibration-star slope, $s$, and the \nuo-ratio errors:

\beq
\lab{const}
      \delta T = \mid\f{\Delta T}{\Delta r}\mid \times\vep_{\rm r} = 
                              s\times\vep_{\rm r}\ \simeq\ {\rm constant}\ .
\eeq
\noindent
This `constant' averages $10\pm3$~K. We also see there is an apparently 
limiting mimimum error, $\vep\sim0.005$, which is presumably 
set by our methods (resolving power, $S/N$, ratio measurements) and an 
apparently limiting minimum slope, $s$, which is at about $-4$~K for the 
$r\ldr$ distribution. That the relationship in \eql{const} appears to exist is 
very desireable as it implies {\it we are measuring much the same temperature 
variations of \nuo with all our ratios}, as 
in fact ideally we should be. This is would also seem to be an important 
detail for our final temperature-precision claims. 
The \delt values are also given in \tl{ratiostats} and indicate, even at this 
stage in our analysis, individual ratio sensitivities lie between 
\mbox{$7-15$\ K}.

Our 215 \nuo spectra and 24 LDRs provided 5160 temperatures. The corresponding 
errors were estimated using the calibration-star-LDR slope and LDR error: 
$\vep_{_{\rm T}}=\vep_{\rm r}\times\Delt/\Delta r$, 
the average error being $12\pm3$~K. Rather than tabulate the mean temperature 
for each ratio we provide in \tl{ratiostats} their differences from our 
expected value ($\tbv =4858$~K). These differences, all of which indicate an 
underestimation of the temperature relative to \tbv, have a mean of 
$-115\pm33$~K, a similar level of accuracy as found for the calibration stars 
for our LDR treatment. Finally, for each ratio in \tl{ratiostats}, the 
regression-line temperatures for the calibration stars are compared to their 
\tbv values and the standard deviation, $\si$, of their differences included. 
Two ratios, $r4356$ and $r5256$, stand out as relatively poor examples for 
recovering the original temperatures, since the associated standard 
deviations, $\si\ldr$, exceed 
220~K whilst the remainder have a mean $120\pm20$~K (see \tl{ratiostats}).

\su{Improving the regression-line predicted temperatures}
\lab{mldrplot}
The next step is to derive a second linear fit, with y-intercept $a$ and 
slope $b$, this time to the residuals of the LDR-\tbv fit, but now in relation 
to \mdiff. This step brings our second H-R diagram coordinate into play. 
Examples of two such 
distributions are given in \fl{residfits}. These two examples (for $r5246$ and 
$r5247$), which are close to the extremes for scatter, were chosen to 
show two other features: 1.\,the scatter of LDR residuals to \mdiff varies 
considerably and is independent of the wavelength separation of the lines used 
for the ratio, and, 2.\,these and all other distributions have a small and 
positive slope indicating there is a remaining sometimes well-defined but 
incompletely-corrected correlation between the LDR and our 
evolution index. We assume the greater scatter of the distributions for some 
ratios represents their greater sensitivity to effects not yet adjusted for 
(surface gravity, metallicity and so on). The two ratios with the highest 
scatter include the Fe\sci line 6256\AA\ -- $r4356$ and 
$r5256$. The next four highest scatters are for ratios including the Fe\scii 
line 6247\AA. This latter result is consistent with findings commented upon 
by Catalano \oo 
(2002), whose two LDRs including this line ($r4347$ and $r4647$) had such high 
gravity-dependence that they could not be used for all of their calibration 
stars. Similarly, our ratio $r5253$ shows a very small scatter and Catalano 
\oo illustrated the tiny gravity effect this ratio has by the near coincidence 
of their main-sequence and giant star calibration curves.

This second linear fit provides the final step to our temperature calibration 
and is often referred to as a `correction' to the LDR, but we prefer to label 
it as a `modification'. The absolute-magnitude modified LDR, 
which we label MLDR, is given by the simple expression

\beq
\lab{mldr}
     r\mldr = r\ldr - (a + b\,\mdiff)\ .
\eeq
\noindent
The distributions and the linear-regression fits of six of these modified LDRs 
are included in \fl{ldrmldr}. It can be seen that each MLDR line has less 
scatter and also less slope than the respective LDR distributions. The 
improvement to the accuracy of our mean regression-predicted 
temperatures of our 20 calibration stars is significant (see \tl{ldrtemps}). 
Now $\mid <T\mldr> - \tbv\mid\ \sim 46\pm23$~K, which is slightly better than 
half the corresponding value given in \scl{first} from the LDRs, and serves as 
an indication of the final accuracy of the temperature scale we will apply to 
our \nuo analysis.

\bfi
\bc
\rotatebox{-90}{\scalebox{0.3}{\includegraphics{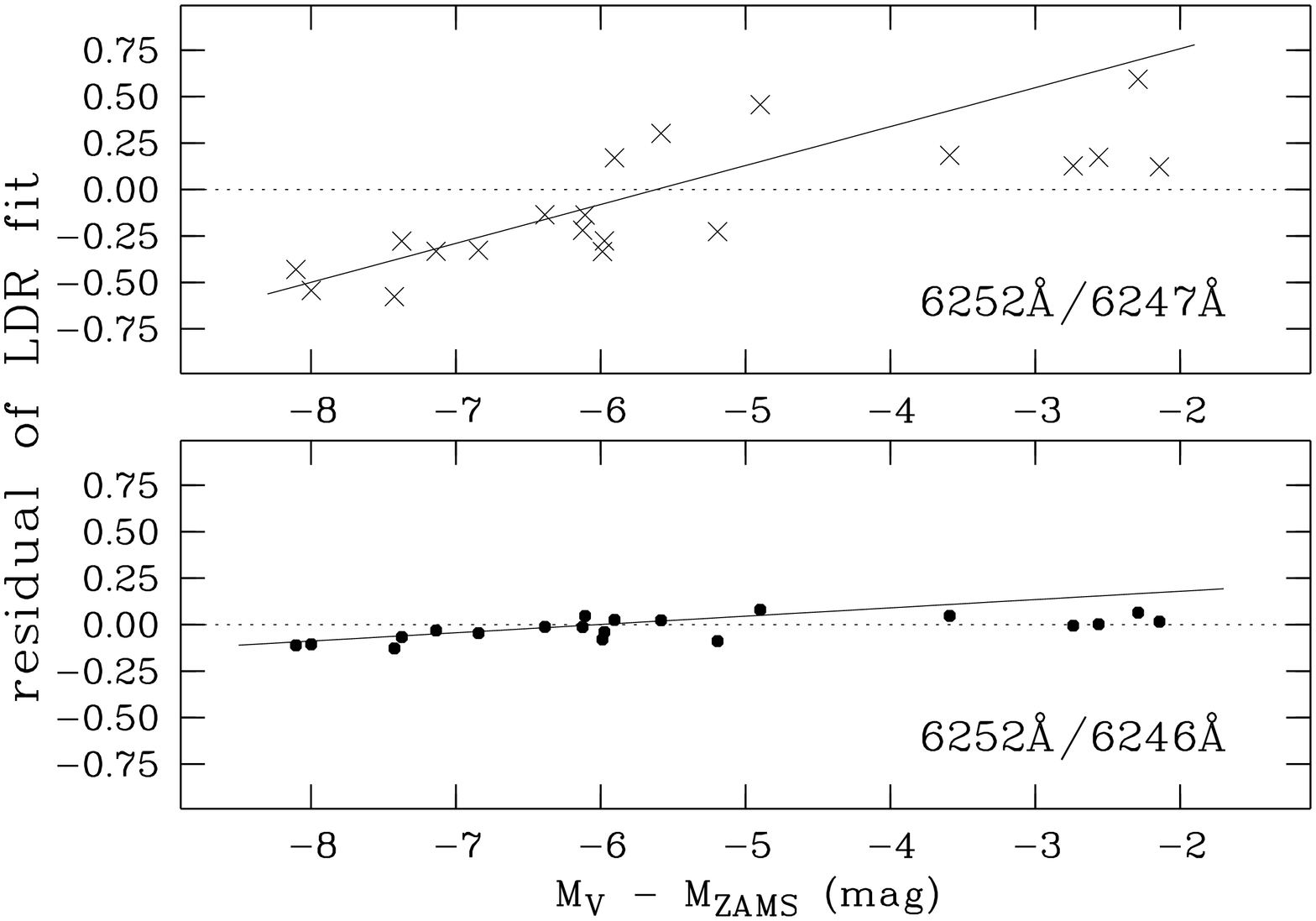}}}
\caption{Two examples of the distribution of residuals of the linear fits to 
\tbv-LDR in relation to our evolution index \mdiff. For each ratio a 
second linear fit was calculated (the sloping solid lines shown here) that 
created the modified ratio (MLDR) used to derive our final temperatures.}
\label{residfits}
\ec
\efi

\su{Final temperatures for $\nu$ Octantis}
The MLDR for each \nuo observation was created using \eql{mldr} and these 
converted via the calibration-star-regression to the 5160 temperatures for our 
215 spectra. The calibration-star-regression slopes, $s$, and temperature 
differences between our \nuo \tbv and $T\mldr$ values, $\nabla T$, are added 
to \tl{ratiostats}. The differences $\nabla T$ are significantly reduced.

Before proceeding it is now appropriate to review all the ratios and decide if 
any can justify rejection. Most of the papers we cite comment upon the 
presence of ratios they deem suitable for rejection initially or as their 
analyses progess. For 
instance, the metallicity dependency of lines and their 
ratios is a function of the degree of their saturation, weak lines having 
almost no metallicity dependence (Gray 1994). Since we have made no adjustment 
for metallicity, we might expect our ratios using a more saturated line (such 
as at 6253\AA ) might yield poorer accuracy and so warrant rejection. This 
dependency is not evident from 
our results and helped us decide not to include a metallicity adjustment for 
our LDRs.\fn{Another reason was the considerable variation of metallicity 
values for any star apparent from a review of values in such databases as 
VizieR. Indeed, \nuo has had its [Fe/H] recently revised from $-0.11$ (Eggen 
1993) to $+0.18$ by Fuhrmann \& Chini (2012), a difference that is not at all 
unusual between different studies.} However, two ratios, $r4356$ and $r5256$, 
were identified in 
our LDR analysis as having the least accuracy recovering the original 
\tbv-calibrated temperatures. This 
weakness continues with the MLDRs. The other 22 ratios have a fairly tight 
mean standard deviation for the calibration-star temperatures of $59\pm7$~K, 
whereas $\si\mldr>110$~K, or at least 7\si, for these two less accurate 
ratios (see \tl{ratiostats}). They are therefore discarded from the final 
analysis.\fn{These are the only ratios that have 
$\nabla\mldr = T\mldr - \tbv > 0$ showing they appear to make the star too hot 
from this pair. Gray (1994) also comments on this related detail.}

As the $\nabla\mldr$ values in \tl{ratiostats} demonstrate, there are 
zero-point offsets between the regression-line temperatures ($T\mldr$, \ie 
using the solid lines as in \fl{ldrmldr}) from our 
remaining 22 ratios ($\si=27$~K). The mean temperature for each \nuo 
observation from these ratios was calculated without any arbitrary 
zero-point adjustment. This decision is the only sensible one given our 
goal to assess the likely true variability of our dataset. Adjusting for the 
offsets can be expected to only unfairly reduce the scatter. The mean 
temperature from the 22 ratios, $T_{\nu Oct}=4810\pm27$~K, is 
consistent with the published effective temperature (Fuhrmann \& Chini 2012) 
who found $\teff=4860\pm40$~K. It is also consistent with the accuracy of 
our mean regression-predicted temperatures of our 20 calibration stars, 
$46\pm23$~K, as mentioned above. The MLDR-calibrated temperatures for the 215 
\nuo spectra, now taking the weighted mean of the 22 ratio temperatures for 
each spectrum, are illustrated in \fl{nuoctteffs}. Their standard deviation is 
only 4.2~K. 
Not surprisingly, given the lack of support for any significant periodicities 
in our original LDRs (see \scl{rotation}), once again a Lomb-Scargle 
periodogram reveals no significant power above the general noise at any 
period.

\bfi
\bc
\rotatebox{-90}{\scalebox{0.3}{\includegraphics{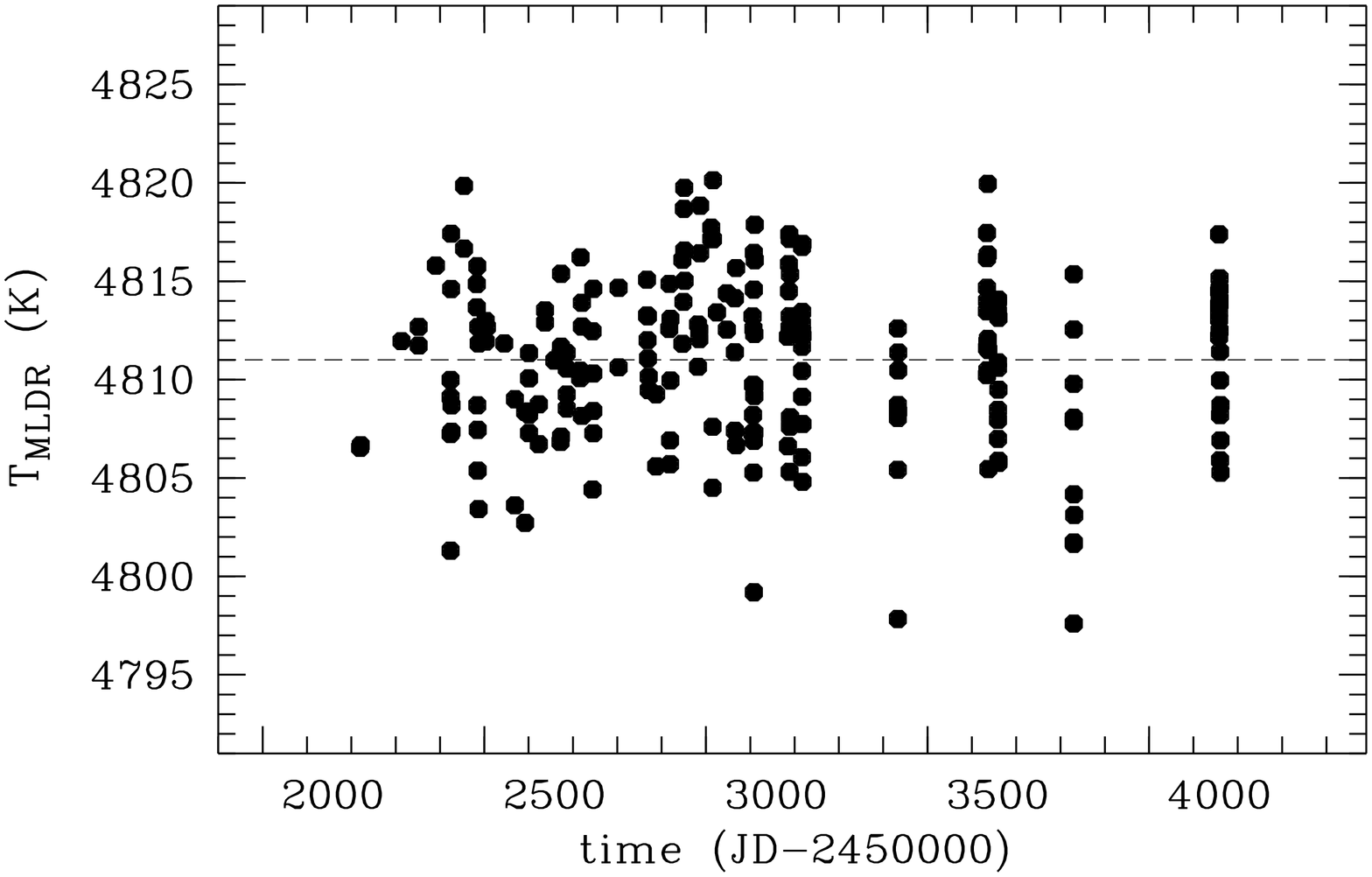}}}
\caption{The final MLDR-calibrated temperatures for $\nu$ Octantis. The mean 
is $4811.0\pm4.2$~K.}
\label{nuoctteffs}
\ec
\efi

\se{Discussion}
Our results so far provide no evidence for significant surface temperature 
variations that are presumably essential evidence of either spots or pulsation 
to be the cause of the $\sim400$~day RV-perturbation of $\nu$ Octantis. 
Another diagram provides 
further evidence, this time pushing our claims back to the observations of 
\hip (ESA 1997). If we assume \nuo is not pulsating \ie $\Delta \mc{R}=0$, we 
can derive the corresponding magnitude variations, $\Delta m$, using 
$M_{\rm bol_2}-M_{\rm bol_1}=-2.5\log(L_2/L_1)$. We create ratios of all 
our MLDR temperatures with their weighted mean \ie $T\mldr/T_{\nu Oct}$. Since 
the bolometric correction can be sensibly assumed to be constant for our tiny 
temperature variations, and the challenging decision of how to define the 
stellar radius disappears with the ratio, returning to the Stefan-Boltzmann 
law, $L_2/L_1=(R_2/R_1)^2(T_2/T_1)^4$, we have

\beq
\lab{sblaw}
        \Delta m = -10\ \log\left(\f{T\mldr}{T_{\nu Oct}}\right)\ .
\eeq
\noindent
These values are plotted in \fl{mmags} together with the $\Delta m$ values 
derived from the \hip observations, $\Delta m = H_{\rm obs} - H_{\rm p}$, 
where $H_{\rm p}$ is given in \tl{stellar}, and both datasets are shifted to 
the same zero-point $\Delta m=0$. The distribution of 
MLDR-predicted magnitudes of course duplicates the temperature distribution in 
\fl{nuoctteffs} (though now inverted as a cooler temperature corresponds 
to a more positive magnitude). The striking similarity of our $\Delta m$ 
magnitudes to the \hip magnitudes implies the primary star of \nuo has the 
same distribution of brightness variations as it did $\sim15$~years 
previously. This useful result further justifies extending our analysis from 
one of only LDRs to the more complex one that has converted the LDRs to 
temperatures. The 
brightness variation is very sensitive to the temperature ratio: increasing 
the temperature range by one degree increases the magnitude range by about one 
millimag, so even increasing $\Delta T$ by a mere 10~K would make a 
significant change to our $\Delta m$ distribution. Such a close match for this 
high-precision behaviour seems highly 
unlikely to be coincidental -- presumably the star is behaving the same way in 
2001--2007 as it did from 1990--1993. 

\bfi
\bc
\rotatebox{-90}{\scalebox{0.3}{\includegraphics{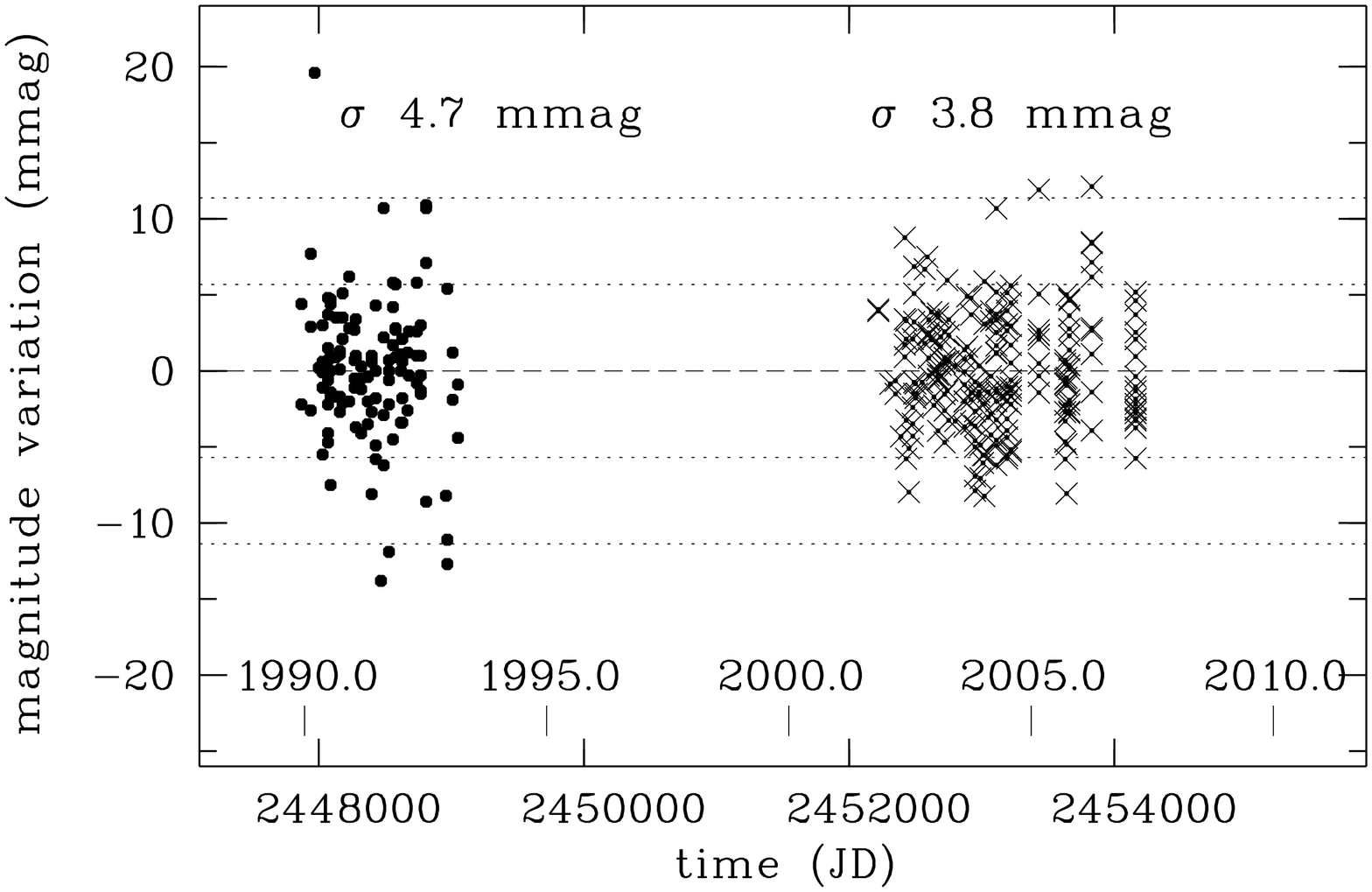}}}
\caption{The implied brightness variation, $\Delta m$, of \nuo based on our 
MLDR-calibrated 
temperature variations `$\times$' compared to the brightness variations 
recorded by \hip `$\bullet$' (only the 116 best-quality $\rm flag=0$ 
observations 
included). The dotted lines identify the $\pm1.5\si$ and $\pm3\si$ limits from 
our MLDR predictions.}
\label{mmags}
\ec
\efi

\su{Pulsations}
Variability amongst K-giants is well recognized, with many having photometric 
and RV periodicities of several hundred days and RV semi-amplitudes up to and 
sometimes in excess of 100\ms (Walker \oo 1989; Cummings \oo 1999; Henry \oo 
2000). The RV period can be comparable to the K-giant rotation period and this 
detail alone should sound alarm bells for the controversial 
\nuo planet. \drb noted this detail, but also that the $(B-V)-\mbv$ 
position of \nuo places it neatly in the space of more typically `RV-stable' 
K giants ($\si_{_{\rm RV}}\lesssim 20\ms$; see \eg Henry \oo 2000, Hekker \oo 
2006). Here we investigate the likelihood of pulsations based on our LDR 
results.

By integrating the near-sinusoidal perturbation RV curve we can 
estimate the change to the star's radius based on the semi-amplitude \krv and 
period $P=400$~days. A highly eccentric orbital solution argues in favour 
of a planetary cause (see \eg the case for the K2~III $\iota$~Dra; Frink \oo 
2002), but 
the small non-zero eccentricity calculated for the conjectured \nuo planet 
($e\sim0.1$) is no obstacle. Also, \drb have assumed their RVs fit a 
Keplerian 
orbit (which in any case may be a significant over-simplification), so it is 
not surprising that some non-zero eccentricity has been derived. We assume 
$e=0$ and therefore $\Delta \mc{R}=\kpul P/2\pi=\pc\krv P/2\pi$, 
where $\pc=\kpul/\krv=1.39$ is a projection correction factor for estimating 
the pulsation velocity amplitude \kpul (Nardetto \oo 2006). For $K$ in \ms, 
$\Delta\mc{R}\sim0.011\krv$\rsun. So for every increment of 10\ms with \krv, 
the predicted change to the star's radius is about 0.1\rsun which acting alone 
corresponds to a brightness change of about 40~mmag. This is far in excess of 
any compensating 
contribution our small temperature variations could provide (\ie merely a few 
millimag). Indeed, the necessary temperature change to offset the brightness 
change with $\Delta \mc{R}$ is about 40~K for every 10\ms increase in \krv. 
These temperature changes would have 
to be in-phase with the radius changes (otherwise the brightness would be more 
variable), which itself is inconsistent with recognized classical pulsation 
behaviour where temperature changes typically significantly lag velocity 
changes (see \eg Kurtz 2006; Pugh \& Gray 2013). Thus, there is no support for 
conventional pulsations as the cause of the \nuo RV perturbation from 
our LDR analyses.

\su{Starspots}
For starspots to cause the RV perturbation, the primary 
star would have to have those spots suitably distributed on the 
observer-facing pole, one spot group if the rotation period \prot matched the 
perturbation period $\ppert\sim400$~days, and two groups fortuitously 
separated by 180\dg if $\prot = 2\ppert$, as \drb noted. Such `fortuitous' 
spot geometries have been identified in the active longitudes of other 
stars (see \eg J{\"a}rvinen \oo 2005; Gray \& Brown 2006). In any case, in 
\scl{rotation} we found $\prot\sim140$~days, and its published radius 
and $v\sin i$ (see \tl{stellar}), at least imply \prot is not likely to 
exceed 200~days.

But it is again not clear that our temperature variations, if now from spots, 
are capable of producing the RV perturbation unless we convert the 
temperatures to some estimate of spot geometry and subsequently RV behaviour. 
We define the filling factor 
$f$(\%) as the ratio of the total spot area $A_s$ to the visible hemisphere 
area $A_v$. We can derive various equations in terms of $f=100\times A_s/A_v$ 
and the unspotted photosphere temperature (assumed to be the maximum observed) 
$T_{\rm ph}$, the integrated observed mean temperature $T_{\rm obs}$ 
when cooler spots are on the visible hemisphere, and the spot temperature 
$T_{\rm s}$. The 
difference between $T_{\rm obs}$ and $T_{\rm ph}$ corresponds to the 
brightness variation $\Delta m$. O'Neal (2006) discusses the difficulties 
inherent in calculating a `meaningful' average temperature when spots 
are involved. In any case, we have our empirical temperatures $T_{\rm obs}$ so 
calculating 
them is unnecessary. O'Neal also emphasises concerns for using LDRs to measure 
spot temperatures $\lesssim4000$~K. Our analyses so far, and what now follows, 
suggests these particular concerns are not relevant here.

We estimate the filling factor 
in a relatively simple manner that should give sensible order-of-magnitude 
predictions of what we are now exploring. In the first place, the ratio of the 
spotted and unspotted luminosities, $L_2$ and $L_1$ respectively, is 
given by

\bea
      \f{L_2}{L_1} &=& 
             \f{(A_v - A_{\rm s})T_{\rm ph}^4 + A_sT_{\rm s}^4}
	       {A_v T_{\rm ph}^4 }\nonumber\\
  &=& 1 - \f{f}{100}\left[1-\left(\f{T_{\rm s}}{T_{\rm ph}}\right)^4\right]\ .
\eea
\noindent
By application of the Stefan-Boltzmann law (since again we assume 
$\Delta \mc{R}=0$) we can then derive 

\beq
\lab{sblawb}
         \f{f}{100} = \f{1-10^{-0.4\Delta m}}{1-(T_s/T_{\rm ph})^4}
\eeq
and
\beq
\lab{tobs}
        \f{f}{100}  =  \f{T_{\rm ph}^4 - T_{\rm obs}^4}
                               {T_{\rm ph}^4 - T_s^4}
\eeq
\noindent
since $L_2/L_1$ also equals $(T_{\rm obs}/T_{\rm ph})^4$. \eql{tobs} allows 
estimation of $f$ when empirical temperatures are available such 
as we have from our MLDR calibrations, and would show the filling factor 
changing in a non-periodic manner consistent with our temperature distribution.

\eql{sblawb} is 
useful for estimating maximum $f$ when a given brightness constraint 
exists, such as we have demonstrated in \fl{mmags}. The standard deviation of 
the brightness variations from \hip (which cannot be seriously questioned) and 
our MLDRs (which are based on very precise temperatures and the sensible 
alternative $\Delta \mc{R}=0$) are both 
$\Delta m\sim 4$~mmag which bounds about 70 per cent of our MLDR datapoints. 
Extending the boundary to $1.5\Delta m\sim 6$~mmag includes 90 per cent 
of them. Both limits are included in \fl{spots}, which illustrates the 
variation of $f$ for three values of $\Delta m$, a wide 
range of spot temperatures, and two photosphere temperatures $T_{\rm ph}$ 
including the maximum MLDR-calibrated temperature we derived for \nuo, 4820~K.
These two temperature curves demonstrate the influence of $T_{\rm ph}$. Each 
value for $\Delta m$ plotted in \fl{mmags} creates its corresponding estimate 
for $f$. At these very low values for $\Delta m$, $f$ is not very senstitive 
to $T_{\rm ph}$.

With regards our plan to estimate spot-induced RV predictions, 
we utilise the work of Hatzes (2002), who also used a photosphere temperature 
of 6000~K. However, the Hatzes analysis was specifically for cool spots on 
sun-like stars. Thus, spottedness of a different nature cannot be reliably 
assessed in what follows. For $-1400<\Delt <-800$~K, 
expected to be typical for many spots (see \eg Biazzo \oo 2006; O'Neal 
2006), each filling factor curve varies very little, and less so for greater 
$\mid \Delt\mid$. For any photosphere temperature 
$4800\leq T_{\rm ph}\leq 6000$~K and $\Delta m\le6$~mmag, we find $f$ does 
not exceed about 1\%. The RV-amplitude prediction from Hatzes (2002) for 
this low filling factor and $v\sin i=2\kms$ is $\lesssim15\ms$. Also, \krv is 
approximately proportional to $v\sin i$ for our model values, so an 
unrealistic increase to $v\sin i$ would be required to 
approach the order of magnitude of $K_{\rm pert}$. That this signal is not 
evident in our temperature 
distribution, and hence neither in our predicted $f$ distribution (there 
is no hint of the required rotation-modulated periodicity), and since our 
predicted spots seem incapable of producing the observed RV perturbation, 
supports -- now quantitatively with data from the same time interval -- the 
claim of \drb that cool spots are unlikely to be the perturbation's cause.

\bfi
\bc
\rotatebox{-90}{\scalebox{0.3}{\includegraphics{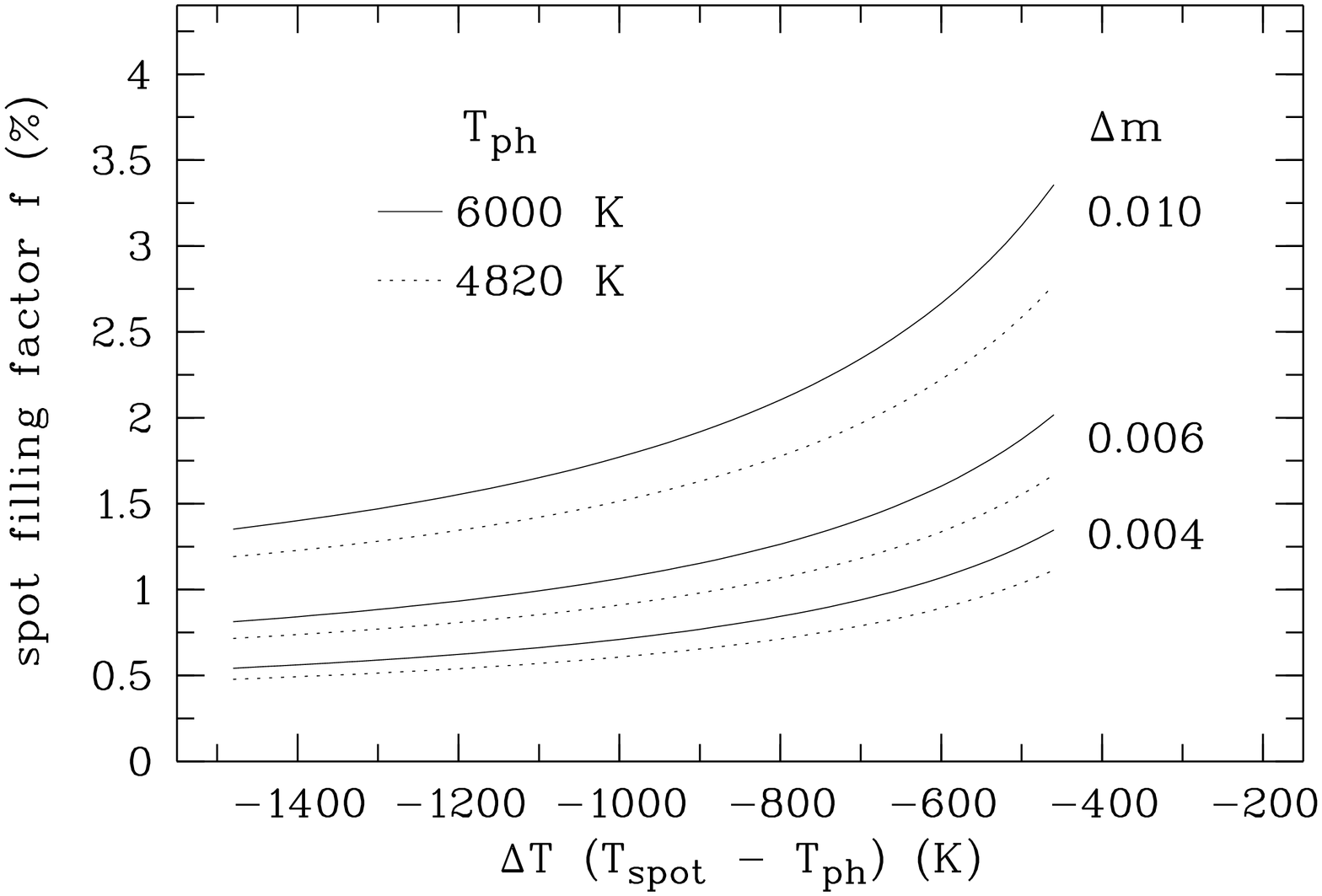}}}
\caption{The predicted spot filling factors created by \eql{sblawb} 
corresponding to two unspotted photosphere temperatures $T_{\rm ph}$ and three 
brightness variations $\Delta m$~(mag).}
\label{spots}
\ec
\efi

\se{Conclusions}
This paper confirms the extraordinary temperature precision achievable from 
line-depth ratio analyses, in this instance for the K0~III/IV primary star in 
the SB1 $\nu$ Octantis. An important consequence is that the conjectured 
circumstellar retrograde planet of this unusually tightly bound system has 
further support. If it wasn't for the close stellar companion, it would be 
unlikely that the \nuo planet would be in question given the range of standard 
tests so far applied and passed, and which now includes the sensitive but 
less frequently used LDR strategy.

Twenty similarly-evolved calibration stars and 24 ratios from high-resolution 
spectra were examined resulting in a final 
set of 22 ratios providing very consistent results.  When compared to 
temperatures derived from the intrinsic colours, 
our MLDR-calibrated temperatures for both our calibration stars and \nuo, 
recover our original \tbv temperatures typically to within about 45~K. 
Our final 215 temperatures for \nuo, spanning several years of 
observations (2001--2007), predict $T=4811$~K with a standard deviation of 
only 4.2~K. The published effective temperature is $4860\pm40$~K (Fuhrmann \& 
Chini 2012).

These results provide the first quantitative evidence that the primary star in 
$\nu$ Octantis has no significant temperature variability during the six-year 
observation window. In particular, there is no evidence 
for any significant periodic behaviour in the LDRs anywhere near the star's 
expected rotation period $\prot\sim140$~days nor near the RV-perturbation's 
period, $\ppert\sim400$~days, the latter suspected of being caused by a 
retrograde planet. When converted to brightness variations, our temperatures 
imply the star is unchanged since its very low-variablity observations by 
\hip, $\sim15$~years previously (ESA 1997). Thus we have also demonstrated a 
novel and revealing use of LDRs as a photometric gauge.

The ability of LDRs to examine a 
fundamental property such as temperature and its many ramifications, as we 
have initiated for the case of \nuo, should serve as a reminder that LDRs 
presumably deserve more widespread use to help 
elucidate the true nature of other exoplanet candidates, perhaps particularly 
but certainly not limited to those that are also controversial, challenged or 
extraordinary in some way.

These results more strongly support the conclusions of Ramm \oo that 
conventional spots and pulsations are unlikely to be the cause of \nuo's 
RV-perturbation. A prograde orbit has no stability, and the binary-secondary 
scenario proposed by Morais \& Correia (2012) appears to be unsupported by the 
available orbital solutions. 
Therefore the only recognizable astrophysical scenario continuing 
to be consistent with all available data is a retrograde planet 
(Ramm \oo 2009;  Eberle \& Cuntz (2010). Thus, the reality of a planet in this 
unusually tight binary system's 
geometry ($\abin<3$~AU, $\apl/\abin\sim0.5$) has more credibility and, for the 
time being, becomes that much more controversial.

Our study appears to be sufficiently robust that it would seem difficult to 
explain the star's temperature stability if in the future surface dynamics 
were instead identified as the cause of \nuo's RV behaviour. Such a 
result would imply that the RV behaviour is caused by a similarly unexpected 
cause, namely a new type of RV-creating surface process that has, in terms of 
our present knowledge, conflicting characteristics - namely be able to create 
a significant RV signal without any photometric, LDR nor bisector evidence. 

Of course new surface phenomena can be imagined to properly explain new 
empirical evidence. Such an example of a variation of otherwise common 
stellar surface features (namely sunpots and starspots) are 
the low-contrast `starpatches' proposed by Toner \& Gray (1988). Similarly, 
Hatzes \& Cochran (2000) investigated the possibility of 'macroturbulent' 
spots for the behaviour of Polaris (where the spot was distinguished by having 
a substantially lower macroturbulent velocity than the surrounding surface). 
However, predictions of such new 
phenomena have their own risks, since these can in turn be erroneous. For 
instance, in a detailed analysis of 51-Peg spectra ($R\sim100,\!000$) which 
included LDRs and bisectors, Gray (1997) and Gray \& Hatzes (1997) proposed 
that a better alternative to that planet (Mayor \& Queloz 1995) seemed to be 
a new mode of stellar oscillation in solar-type stars. However, using 
$R\sim220,\!000$ spectra, Hatzes \oo (1998) were able to help confirm the 
now-accepted reality of the planet. This should provide a warning 
for the challenges inherent in using even spectra with $R$ as high as 
\mbox{100,000} for bisector and LDR analyses. Of course, a large fraction of 
planets 
have been both supposedly confirmed and others refuted with spectra of only 
modest resolving power. For instance, HD\,166435's planet was refuted in a 
frequently cited paper by Queloz \oo (2001) with spectra having only 
$R\sim42,\!000$. Indeed it remains unclear just what minimum resolving power 
is reliable 
for bisector and LDR studies in each instance, and, specifically, it remains 
to be proven if the spectra used by Ramm \oo 
(2009) for their bisector analysis, and again used here ($R\sim70,\!000$) are 
in fact truly adequate to reveal the tell-tale evidence of a non-planetary 
cause. This detail perhaps remains the biggest obstacle for greater confidence 
that the extraordinary \nuo planet is real.

If proposals of new surface behaviour are possible for solar-type stars, it is 
surely possible for evolved stars about which we have less certain knowledge. 
However, if our increasing list of reasons to support the reality of the \nuo 
planet is confirmed, such as of the lack of suitable bisector and 
LDR variability with higher resolving-power spectra, should the controversial 
planet be later disproven by the discovery of another cause, it 
would presumably have serious implications for many exoplanet claims.

The challenges the \nuo system presents for its formation are formidable and 
challenging for long-term stability theories as well. But unless 
somehow discredited in the future, the \nuo system will be a prime 
motivator for studying such demanding geometries. Besides being consistent 
with the debris-disk proposals in similar close binaries as mentioned in our 
Introduction (\ie by Trilling \oo 2007), perhaps 
the \nuo planet will be a suitable candidate for such histories as 
`star-hopping', whereby a planet in a binary, 
rather than being ejected by collisions from a passing star or strong 
interactions from close stellar companions or an evolving stellar host, 
instead is exchanged between the stars (see \eg Kratter \& Perets 2012). 
Such an exchange may explain the proposed retrograde orbit.

\section*{Acknowledgments}
The author thanks the referee, A.~Hatzes, for his reading of the paper and 
his helpful comments. 
The author is grateful for the availability of 18 spectra acquired by 
S.~Komonjinda, J.~Skuljan and J.~Hearnshaw. This work benefited from NASA's 
Astrophysics Data System, the SIMBAD database and the VizieR catalogue access 
tool, CDS, Strasbourg, France, and the Vienna Atomic Line Database (VALD3).

\end{document}